\RequirePackage{fix-cm}

\documentclass[smallextended]{svjour3}       
\smartqed  
\usepackage{graphicx}
\usepackage{multirow}
\usepackage{color}
\usepackage{listings}
\usepackage{float}
\usepackage{arcs}
\usepackage{aas_macros}
\usepackage{natbib}
\usepackage{caption}
\usepackage{subcaption}
\usepackage{upgreek}
\usepackage{wasysym}
\usepackage{url}
\usepackage{mathptmx}      

\usepackage{listings}
\usepackage{newfloat}

\definecolor{light-gray}{gray}{0.95} 

\DeclareFloatingEnvironment[name=Listing]{python}
\DeclareSymbolFont{yhlargesymbols}{OMX}{yhex}{m}{n}
\DeclareMathAccent{\wideparen}{\mathord}{yhlargesymbols}{"F3}

\newcommand{\unit}[1]{\textnormal{#1}}

\def\arcsec{\hbox{$^{\prime\prime}$}}
\def\degr{\hbox{$^\circ$}}
\def\arcmin{\hbox{$^\prime$}}
\def\fdg{\hbox{$.\!\!^\circ$}}
\def\farcm{\hbox{$.\mkern-4mu^\prime$}}
\def\farcs{\hbox{$.\!\!^{\prime\prime}$}}
\newcommand{\micron}[0]{$\upmu\unit{m}$}
\newcommand{\htmid}[0]{\texttt{HTMID}}
\newcommand{\nicearc}[1]{\wideparen{\small{#1}}}

\begin{document}

\title{The Footprint Database and Web Services of the Herschel Space Observatory\thanks{The research was supported by the Hungarian Scientific Research Fund OTKA NN 103244 and OTKA NN 114560 grants and PECS contract 4000109997/13/NL/KML of the ESA and the Hungarian Space office. {\it Herschel} is an ESA space observatory with science instruments provided by European-led Principal Investigator consortia and with important participation from NASA.}}


\author{L\'aszl\'o Dobos \and
        Erika~Varga-Vereb\'elyi \and
        Eva~Verdugo \and
        David~Teyssier \and
        Katrina~Exter \and
        Ivan~Valtchanov \and
        Tam\'as~Budav\'ari \and
        Csaba~Kiss
}


\institute{L.~Dobos \at
              Department of Physics of Complex Systems, E\"otv\"os Lor\'and University, \\
              P\'azm\'any P\'eter s\'et\'any 1/A, 1117 Budapest, Hungary \\
              \email{dobos@complex.elte.hu}
           \and
			  E.~Varga-Vereb\'elyi, C.~Kiss \at
              Konkoly Observatory, Research Centre for Astronomy and Earth Sciences \\
              Hungarian Academy of Sciences \\
			  H-1121 Budapest, Konkoly Thege Mikl\'os \'ut 15-17, Hungary  \\
              \email{verebelyi.erika@csfk.mta.hu, pkisscs@konkoly.hu} 
		   \and
		      E.~Verdugo, D.~Teyssier, K.~Exter, I.~Valtchanov \at
		      European Space Astronomy Centre, European Space Agency \\
		      Villanueva de la Ca\~nada, 28691 Madrid, Spain \\
		      \email{everdugo@sciops.esa.int, dteyssier@sciops.esa.int, kexter@sciops.esa.int, ivan.valtchanov@sciops.esa.int}
           \and
              T.~Budav\'ari \at
              Department of Applied Mathematics and Statistics, The Johns Hopkins University \\
              3400 North Charles Street, Baltimore, Maryland 21218, USA \\
              \email{budavari@jhu.edu}
}

\date{Received: \today / Accepted: date}


\maketitle

\begin{abstract}
Data from the Herschel Space Observatory is freely available to the public but no uniformly processed catalogue of the observations has been published so far. To date, the Herschel Science Archive does not contain the exact sky coverage (footprint) of individual observations and supports search for measurements based on bounding circles only. Drawing on previous experience in implementing footprint databases, we built the Herschel Footprint Database and Web Services for the Herschel Space Observatory to provide efficient search capabilities for typical astronomical queries. The database was designed with the following main goals in mind: (a) provide a unified data model for meta-data of all instruments and observational modes, (b) quickly find observations covering a selected object and its neighbourhood, (c) quickly find every observation in a larger area of the sky, (d) allow for finding solar system objects crossing observation fields. As a first step, we developed a unified data model of observations of all three Herschel instruments for all pointing and instrument modes. Then, using telescope pointing information and observational meta-data, we compiled a database of footprints. As opposed to methods using pixellation of the sphere, we represent sky coverage in an exact geometric form allowing for precise area calculations. For easier handling of Herschel observation footprints with rather complex shapes, two algorithms were implemented to reduce the outline. Furthermore, a new visualisation tool to plot footprints with various spherical projections was developed. Indexing of the footprints using Hierarchical Triangular Mesh makes it possible to quickly find observations based on sky coverage, time and meta-data. The database is accessible via a web site\footnote{\url{http://herschel.vo.elte.hu}} and also as a set of REST web service functions, which makes it readily usable from programming environments such as Python or IDL. The web service allows downloading footprint data in various formats including Virtual Observatory standards.


\end{abstract}

\section{Introduction}
\label{sec:intro}

One of the most frequent tasks in astronomy is to collect information on a specific source obtained by different instruments, often at different wavelengths. For this purpose, any astronomical database must support efficient search by coordinates and contain information on the sky coverage. Knowing the exact sky coverage is indispensable for most statistical studies such as identifying members or clusters of stars and galaxies, and it is also useful for studying the structure of the interstellar medium at multiple wavelengths. Existing databases that already provide exact sky coverage include large homogeneous surveys such as the Sloan Digital Sky Survey \citep{Budavari2007, Budavari2010} and observatory archives such as the Hubble Source Catalog \citep{Lubow2013}.

The Herschel Space Observatory \citep{Herschel} was a cornerstone mission of the European Space Agency, operating between 2009 and 2013. The three instruments of the telescope provided imaging and spectroscopic observations at far infrared and submillimetre wavelengths (55 -- 672\,\micron{}) to study the formation of galaxies in the early Universe and their subsequent evolution and to investigate the birth of stars and their interaction with the interstellar medium.

All observations performed by Herschel are stored in the Herschel Science Archive (HSA) and accessible by the astronomy community through the HSA User Interface\footnote{\url{http://archives.esac.esa.int/hsa}} or via the Herschel Interactive Processing Environment (HIPE, \citealp{Ott2010}). The HSA has been constantly updated with improved reductions of the observations. To this date, however, the HSA does not contain the exact description of observation footprints and lacks the functionality to perform fast coordinate-based spatial search on footprints. With this work we attempt to bridge this gap by building a database of unified observation headers and footprints. Photometric footprints presented here are already part of the HSA since version 7.0 and spectroscopic footprints will be added soon. The database was created with the following in mind:
\begin{itemize}
\item provide uniform access to observations made by all three instruments
\item to find every observation for a selected object and its neighbourhood
\item to find every observation in a larger area in the sky
\item to find Solar System objects intersecting with Herschel footprints and plot their trajectories on the map
\end{itemize}
The database can further be extended to include a source catalogue and become an integral part of the HSA, the Virtual Observatory and other astronomical data services such as SkyQuery \citep{SkyQuery} or Spectrum Services \citep{Dobos2004}.

In Sec.~\ref{sec:herschel} we summarise the operation modes of Herschel and the data reduction system for reference. Sec.~\ref{sec:footprints} introduces the features of the Herschel Footprint Database including the user interface and web services. The building process of the database is explained in Sec.~\ref{sec:building}. For the more database-inclined readers, we review the methods of representing and indexing spherical geometries in relation databases in Sec.~\ref{sec:regions}-\ref{sec:htm}.  We present the two outline reduction methods that we implemented to simplify the complex footprints of scan maps in Sec.~\ref{sec:reduce}. The paper is summarised in Sec.~\ref{sec:summary}.

This paper is intended for the end users of the footprint database; a technical paper addressing the issues regarding fast search among footprints is in preparation by Dobos~et~al. (2016).

\section{The Herschel Space Observatory and its observation data}
\label{sec:herschel}

Herschel was deployed at the L2 Lagrangian point of the Earth-Sun system in 2009 and was in operation until its He coolant depleted in 2013. During this period, Herschel's instruments took about $46{,}000$ observations (including calibration and failed) of solar system, galactic and extragalactic sources in the far infrared and sub-millimetre. Many observations were focused on the photometry or spectroscopy of specific point sources, thus covered very small areas. In addition, there are about $1{,}500$ raster maps and almost $28{,}000$ photometric scan map observations, of which about $5{,}300$ covered an area larger than $0.1$~sq.deg.

The Herschel spacecraft featured a $3.5$~metre warm mirror($80$-$90$~K) with a field of view of approximately $15$\arcmin. Instruments were cooled below $2$~K by the cryogenic system. Tab.~\ref{tab:photometers}-\ref{tab:spectroscopy} summarises the most important properties of Herschel's instruments. The guiding system of the observatory allowed for very precise movement and pointing of the telescope which resulted in an astrometric accuracy of about 2\arcsec.

\subsection{Pointing modes}
\label{sec:pointingmodes}

To suit differential measurements and to create maps much larger than the field of view of the instruments, Herschel supported a wide variety of pointing modes. The astronomical observation templates offered four fundamental pointing modes: single pointing, raster maps, nodding and scanning. In addition to positioning the telescope, instruments had the ability to manipulate their field of view by means of a chopper or beam-steering mirror. Moreover, the telescope was able to track moving solar system objects along a pre-programmed trajectory (tracking mode).

\paragraph{Single pointing}: The telescope was positioned on a certain coordinate with high accuracy and differential measurement were made using the the chopper or beam-steering mirror of the instruments. Certain observation templates extended the simple pointing with \textit{jiggling}, moving the boresight of the telescope slightly around a given point for higher spatial sampling, but this mode was not used for scientific purposes.

\paragraph{Nodding}: To take differential measurements, the telescope could alternate between a target position and off-position calibration fields. This mode could be combined with chopping in case of certain observation templates.

\paragraph{Raster maps}: The telescope was able to cover extended areas by combining multiple, equally spaced single pointings. This mode was primarily used for integral field spectroscopy.

\paragraph{Scanning}: All scientific imaging observations were taken in scanning mode. Scan maps were observed by combining constant velocity parallel and almost perpendicular line scans into a single photometric observation to cover extended sources. 

\subsection{The PACS instrument}

The shorter-wavelength instrument capable of imaging photometry and integral field spectroscopy was the Photodetector Array Camera and Spectrometer (PACS, \citealp{PACS}, \citealp{PACSHandbook}).

\subsubsection{The PACS photometer}

Inside the PACS photometer the light beam from the telescope was split by a dichroic making parallel observations of the same field possible by the two bolometer matrices. The longer wavelength detector had a resolution of $32 \times 16$ pixels and a fixed broad-band filter with a coverage between 125 -- 210\,\micron{}, nicknamed \textit{red}. The shorter wavelength detector had $64 \times 32$ pixels and two interchangeable filters covering the wavelength ranges of 60 -- 85\,\micron{} and 85 -- 125\,\micron{}, called \textit{blue} and \textit{green}, respectively. The instantaneous field of view of the imaging detectors was equally $1\farcm75 \times 3\farcm5$. To be able to perform differential measurements without telescope movement, a chopper mirror was also installed inside PACS but it was never used for scientific imaging observations. Instead, scan maps and mini scan maps were created with a great variety of shape, coverage and repetition factor. Three scanning velocities, $10\arcsec$, $20\arcsec$ and $60\arcsec$ per second, were used. As the scanning mode was very inefficient for small maps due to the frequent telescope turn arounds at the ends of the scan legs, only maps of approximately $5'$ in diameter and larger were made with this technique. The resulting coverage was not always homogeneous: PACS scan maps produced by the pipeline include a pixel-wise coverage map, but instead of this pixelised map, it was more convenient to recover the footprints of observations from the telescope pointing data. Whenever it was possible, we utilised actual telescope pointing information, but for certain data products (such as some HIFI raster maps) we had to rely on commanded position angles.

\begin{table}
	\begin{tabular}{l l r r c c}
		\hline\noalign{\smallskip}
		instrument & filter & $\lambda$ coverage & $\lambda$ nominal & field of view$^a$ & pixel scale \\
		\noalign{\smallskip}\hline\noalign{\smallskip}

		\multirow{3}{*}{PACS} &
		blue & 
		$60$ -- $85$\,\micron{} & $70$\,\micron{} &
		\multirow{3}{*}{$1\farcm75 \times 3\farcm5$} & 
		$3\farcs28$  \\

		& green & $85$ -- $130$\,\micron{} & $100$\,\micron{} & & $3\farcs28$ \\
		& red & $130$ -- $210$\,\micron{} & $160$\,\micron{} & & $6\farcs6$ \\
		
		\noalign{\smallskip}\hline\noalign{\smallskip\smallskip\smallskip\smallskip\smallskip}
		
		\hline\noalign{\smallskip}
		instrument & band & $\lambda$ coverage & $\lambda$ nominal & field of view$^a$ & beam FWHM$^b$ \\
		\noalign{\smallskip}\hline\noalign{\smallskip}
		
		\multirow{3}{*}{SPIRE} & PSW & $212$ -- $292$\,\micron{} &
		$250$\,\micron{} & 
		\multirow{3}{*}{$4\arcmin \times 8\arcmin$}	& 
		$17\farcs6$  \\
		
		& PMW & $296$ -- $296$\,\micron{} & $350$\,\micron{} & & $23\farcs9$ \\
		& PLW & $410$ -- $611$\,\micron{} & $500$\,\micron{} & & $35\farcs2$ \\
	
		\noalign{\smallskip}\hline \\
		
		\multicolumn{6}{l}{
			\footnotesize $^a$~~instantaneous field of view
		} \\
		
		\multicolumn{6}{l}{
			\footnotesize $^b$~~as measured on maps with $1''\!/\textnormal{pixel}$ for the nominal scan speed of $30''\!/\textnormal{sec}$
		}
	\end{tabular}
	\caption{Herschel science instruments for photometry.}
	\label{tab:photometers}
\end{table}

\subsubsection{The PACS spectrometer}

The PACS integral field spectrograph had $5 \times 5$ ``spaxels'' organized in an irregular square with a coverage of $47'' \times 47'' $. The instrument could cover selected regions within the spectral ranges $51$ -- $105$ and $103$ -- $220$\,\micron{} simultaneously, either in a line-scan or a range scan mode. About $ 5{,}000 $ PACS spectroscopic measurements were done with single pointing and more than $1{,}000$ raster maps were created. For differential spectroscopic measurements the chopping and nodding techniques were used in combination. An unchopped mode used for crowded fields was also provided.

\begin{table}[ht]
	\begin{tabular}{l l r c c}
		\hline\noalign{\smallskip}
		instrument & band & $\lambda$ coverage & field of view & pixel scale \\
		\noalign{\smallskip}\hline\noalign{\smallskip}
		
		\multirow{2}{*}{PACS} &
		blue &
		$51$ -- $105$\,\micron{} &
		\multirow{2}{*}{$47'' \times 47''$} &
		\multirow{2}{*}{$9\farcs4$} \\
		
		& red & $103$ -- $220$\,\micron{} & & \\
		
		\noalign{\smallskip}\hline\noalign{\smallskip\smallskip\smallskip\smallskip\smallskip}
		
		\hline\noalign{\smallskip}
		instrument & band & $\lambda$ or $\nu$ coverage & field of view & beam FWHM \\
		\noalign{\smallskip}\hline\noalign{\smallskip}
		
		\multirow{2}{*}{SPIRE} &
		SSW & $194$ -- $313$\,\micron{} &
		\multirow{2}{*}{$\diameter 2'$} &
		$29''$ -- $42''$ \\
		
		& SLW & $303$ -- $671$\,\micron{} & & $17''$ -- $21''$ \\
		
		\noalign{\smallskip}\hline\noalign{\smallskip\smallskip\smallskip\smallskip\smallskip}
		
		\hline\noalign{\smallskip}
		instrument & band & $\lambda$ or $\nu$ coverage & beam FWHM$^a$ & max. beam FWHM$^b$ \\
		\noalign{\smallskip}\hline\noalign{\smallskip}
		
		\multirow{7}{*}{HIFI} 
		 & 1 & $480$ -- $640$~GHz & $32\farcs 5$ -- $43\farcs 3$ & $43\farcs 1$ -- $43\farcs 5$ \\
		 & 2 & $640$ -- $800$~GHz & $26\farcs 3$ -- $32\farcs 9$ & $32\farcs 8$ -- $32\farcs 9$ \\
		 & 3 & $800$ -- $960$~GHz & $21\farcs 7$ -- $26\farcs 1$ & $25\farcs 8$ -- $26\farcs 3$ \\
		 & 4 & $960$ -- $1120$~GHz & $18\farcs 7$ -- $21\farcs 8$ & $21\farcs 7$ -- $21\farcs 9$ \\
		 & 5 & $1120$ -- $1250$~GHz & $17\farcs 2$ -- $19\farcs 5$ & $19\farcs 4$ -- $19\farcs 6$ \\
		 & 6 & $1410$ -- $1703$~GHz & $12\farcs 3$ -- $14\farcs 8$ & $14\farcs 7$ -- $14\farcs 9$ \\
		 & 7 & $1703$ -- $1910$~GHz & $11\farcs 1$ -- $12\farcs 4$ & $11\farcs 1$ \\
		 
		
		
		\noalign{\smallskip}\hline \\
		
		\multicolumn{5}{l}{
			\footnotesize $^a$~~frequency dependent, averaged for the two polarisations
		} \\
		
		\multicolumn{5}{l}{
			\footnotesize $^b$~~different for the two polarisations
		}		
		
	\end{tabular}
	\caption{Herschel science instruments for spectroscopy.}
	\label{tab:spectroscopy}
\end{table}

\subsection{The SPIRE instrument}

The Spectral and Photometric Imaging Receiver (SPIRE, \citealp{SPIRE}, \citealp{SPIREHandbook}) was an instrument for far infrared and submillimeter imaging and Fourier-Transform spectroscopy.

\subsubsection{The SPIRE photometer}

The SPIRE Photometer had three hexagonally close-packed feedhorn-coupled bolometer arrays. The incoming radiation from the sky is split into three bands, centred approximately at $250$, $350$ and $500$\,\micron{}. The three arrays observe the same area of the sky simultaneously. The spatial resolution in each band, in terms of the beam FWHM, varies by wavelength as it is listed in Tab.~\ref{tab:photometers}. SPIRE had a beam steering mirror to change the field of view of its detectors with respect to the field of view of the telescopes. It could also be used to take differential measurements against the built-in black-body calibration sources. Although the system originally supported various point source photometry modes (chopping, jiggling, nodding), these were never used for scientific observations. Rather, large and small area scan maps were created by slewing the telescope at constant angular velocity. Large area maps were at least $5\arcmin$ in diameter while small maps were created by scanning two short legs almost perpendicular to each other with lengths shorter than $5\arcmin$. For SPIRE observations three scanning velocities, $20$, $30$ and $60\arcsec/\unit{s}$, were used.

\subsubsection{The SPIRE spectrometer}

The SPIRE imaging Fourier-Transform Spectrometer had two, slightly overlapping bands covering the rather broad wavelength ranges $194$ -- $318$\,\micron{} and $294$ -- $671$\,\micron{} with $\Delta\nu = 1.2$~GHz at high resolution (for spectral lines and continuum) or $\Delta\nu = 25$~GHz at low resolution (continuum) observation. The shorter wavelength detector (SSW) had 19, the longer wavelength detector (SLW) had 37 feedhorn-coupled bolometers closely-packed in a hexagonal pattern, covering the same field of view of $2\farcm6$ in diameter. Some detectors are partially vignetted by the instrument, hence the unvignetted field of view is about $2\arcmin$. Both detectors could be operated simultaneously. The instrument was used for pointed observations and raster maps. Raster maps were made with column separation of $116''$ and row separation of $110''$ to fill in the gaps between adjacent pointings with the hexagonal detectors.

\subsection{The HIFI instrument}

The Heterodyne Instrument for the Far Infrared (HIFI, \citealp{HIFI}, \citealp{HIFIHandbook}) was a spectrograph working from the far infrared to the sub-millimeter band. It had seven beams covering different frequency ranges with diffraction-limited beam diameters between $11\farcs1$ and $43\farcs5$. The electronically tunable mixers and the two pairs of auto-correlator and acousto-optical spectrometers provided a spectral resolution of $125$~kHz-$1$~MHz, at both polarizations in parallel. Precise beam characterisation revealed a small offset between beam centres of the two polarization. The size and direction of the offset varied from band to band and was on the order of arc seconds. For the HIFI instrument a wide range of observation templates were developed with various types of differential measurement techniques. Positional differentiation could be done using the chopper mirror to chop between points on the sky or the calibration source. Spectral differentiation could be done using beam switching or frequency switching. Three different pointing modes were used with HIFI: single pointed, raster maps, on-the-fly scan maps.

\subsection{Herschel observations}

Instead of following a survey schedule, Herschel made measurements based on individual astronomical observational requests proposed by scientists. Each observation is identified by a number, the \texttt{obsID}. Observations are primarily characterised by the instrument(s) used, whether it was broad-band photometry or spectroscopy, the instrument mode and band, and the telescope pointing mode. Herschel's data acquisition system recorded operational data with high resolution in time encoded into the \texttt{fineTime}, a time unit with \micron{} resolution measuring atomic time seconds since midnight 1 January 1958 UT2.  

Herschel operated in numerous observational modes including single pointed observations, raster maps and scan maps. Pointed and raster observational modes were used primarily for spectroscopy while most photometric observations were scan maps. Differential observations in the pointed and raster modes were most often made either by using the instruments' chopper, or nodding with the telescope (or the combination of the two) to alternate between the source of interest and a reference field.

Many observations were made for instrument calibration purposes or failed for any reason. Observations used for calibration purposes are also assigned \texttt{obsID}s and they are usually flagged as `calibration' in the astronomical observation request (AOR) field of the data header by the word 'cal' or 'calib', 'calibration', etc. Certain SPIRE observations were also flagged as 'dark sky' which we also classified as calibration. 

\subsubsection{Herschel photometry}

By original design, the PACS instrument was able to perform differential photometric measurements by chopping between an on and off position on the sky. Although a few observations were made in this mode, cross-scan maps turned out to be superior and were later used for all photometric point source observations.

Scan maps were performed by slewing the boresight of the telescope at a constant angular velocity along parallel lines (scan legs) to cover a large area. The lines followed great circles on the sphere, approximately parallel lines over short distances. After each scan leg, the spacecraft performed a turn-around manoeuvre to continue observing along the next, slightly overlapping scan leg in the opposite direction. The time a turn-around took was about half a minute. Scan maps were taken at $10$, $20$, $30$ or $60\arcsec$ per second angular velocity, depending on the instrument, map size and observational mode used. No differential measurements were made during scan map observations; the instrumental background was removed by high-pass filtering the detector read-out time signal. Larger maps were usually scanned in two orthogonal directions and the whole process was repeated several times for higher signal to noise ratio.

PACS cross-scan maps (also mini scan maps) were the combination of two short ($3'$) scan legs along the detectors' diagonal at angles of $70^\circ$ and $110^\circ$. SPIRE small scan maps were made in a similar way, scanning two short legs almost perpendicular to each other, at angles of $\pm 42.4^\circ$ with respect to the detector to cover a minimum circle of $5'$ in diameter, in the middle of the map.

\subsubsection{PACS/SPIRE parallel photometric scan maps}

The parallel mode was the only compatible mode for operating SPIRE and PACS simultaneously in which the scanning strategy was optimized for both instruments. Using the two detectors of PACS and the three detectors of SPIRE, simultaneous observations in five bands were possible without significant degradation in instrument performance. 
This mode is used to cover very large areas of the sky. For any observation in parallel mode, only scans at one of the two predefined ``magic'' angles (``nominal'': $+ 42.4\degr$ and ``orthogonal'': $- 42.4\degr$ , with respect to the instrument coordinate system) were possible. The scan speed could be either 20 or $60 \arcsec/\unit{s}$ in this mode. Cross-scan maps could also be achieved by additional parallel mode observations at the alternative scan direction. In parallel mode, the PACS and SPIRE sky coverage were offset by $21\arcmin$. The major advantage of PACS/SPIRE parallel mode was for programmes involving shallow Galactic surveys covering dozens of square degrees. When observing in parallel, both PACS and SPIRE observations received the same \texttt{obsID}.

\subsubsection{Spectroscopy}

Herschel's two imaging spectrometers (PACS and SPIRE) were used for single pointing and raster map observations. The single-beam HIFI spectrometer was used for pointed observations, on-the-fly scan maps and raster maps. Because differential measurements were made using various combinations of telescope nodding and chopper movement, recovering the actual area of interest from the raw pointing data was sometimes difficult. Details of the footprints of spectroscopic observations will be highlighted in Sec.~\ref{sec:spec_single} and \ref{sec:spec_mapping}.

\subsubsection{Observation processing levels}

The HSA, via HIPE, provides access to reduced data at various levels of processing. To generate the observation footprints we reached back to the highest observation level data (not necessarily science-ready) which still contained the pointing information. The fully processed data usually contain a WCS header and a coverage mask only which is unsuitable to reconstruct the exact area of interest.

PACS photometry Level-0 and Level-0.5 observations are minimally processed, non-calibrated, averaged images generated from raw bolometer read-outs. At these low processing levels data products still contain the telemetry information necessary to reconstruct the exact coverage of scan maps. Observations at Level-1 processing are fully reduced and flux-calibrated, including the removal of all instrumental and observatory effects, but not ready for scientific analysis. Level-2 scan maps are science-ready and contain, for each camera, the image, the coverage map, the standard deviation, high-pass filtered image and error map. Level-2.5 data is available if multiple scan maps and cross-scans were combined into mosaic maps. Please refer to \cite{Balog2014} for details on the reduction process and photometric calibration of PACS scan maps. PACS spectroscopy Level-0 data consist of the $18 \times 25$ read-out time-lines for each camera, and the calibration data needed for data analysis. At Level-0.5 wavelengths are calibrated and digital read-outs are converted to Volts/s. PACS spectroscopy Level-1 products contain the fully calibrated coordinates of the 5 x 5 spaxels and the flux is calibrated and converted to physical units. Level-2 data products contain noise filtered, regularly sampled spectral data cubes ready for scientific analysis.

SPIRE photometry Level-0 and Level-0.5 are similar to those of PACS but telescope turn arounds during scan map observations are already filtered from the pointing data and are not recoverable. SPIRE photometry Level-1 and Level-2 are essentially the same as for PACS photometry, whereas SPIRE Level-2.5 is the same as SPIRE Level-2.0. SPIRE spectroscopy Level-0 contains raw read-outs from the 19 and 37 detectors. At Level-1, measurements are calibrated and converted to physical units. At Level-2, spectral data cubes are completely assembled.

Level-3 processing is available for PACS and SPIRE if various observations belonging to the same proposal are merged. These are all photometric mosaic images.

HIFI Level-0 data consist of raw read-outs from the detectors, sampling one sub-band of a single mixer with both polarizations at the same time. HIFI Level-1 data contain amplitude and frequency-calibrated measurements corrected for fundamental instrumental effects. At Level-2, additional instrumental effects (e.g. standing waves, baseline offset and slope, side band convolution etc.) are removed. At Level-2.5 processing, spatial maps are assembled into regridded data cubes and sideband deconvolution of spectral scans is performed.

\section{Herschel observation footprints}
\label{sec:footprints}

We downloaded all reduced observations from the Herschel Science Archive, converted header meta data into a uniform format for all instruments and observational modes to build a relation database. As the most important contribution to the Herschel data infrastructure, we computed the exact footprint for all scientific observations taken by the observatory. Footprints were also computed for most observations flagged as calibration or failed where the pointing information was available. Some scientific observations were tracking moving Solar System objects in which case a footprint with absolute coordinates is not meaningful.

\subsection{PACS and SPIRE scan maps}

Scan maps obtained by Herschel cover $15,200$~square degrees. This coverage is comparable with systematic sky surveys and uniformly reduced data can later be published as source catalogues, similarly to the recently created Hubble Source Catalog by \citet*{Lubow2013}. To support the typical survey operations such as spatial searching and statistical analysis, the detailed, spatially indexed description of the footprint of the covered areas is indispensable. Scan maps are usually processed at Level 2.0 or 2.5 which are the equivalent of flux-calibrated mosaic images in FITS format with a mask. Reconstructing the exact footprints of scan maps from these highly processed images would be challenging, mostly due to concave regions. Therefore, we had to reach back to initial pointing data - the pointing history of the telescope in a certain time window - and calculate footprints from the trajectories swept by the field of view of the various instruments.

\begin{figure}

\vspace*{-0.3cm}

\begin{subfigure}[t]{\textwidth}
	\parbox{0.6\linewidth}{\includegraphics{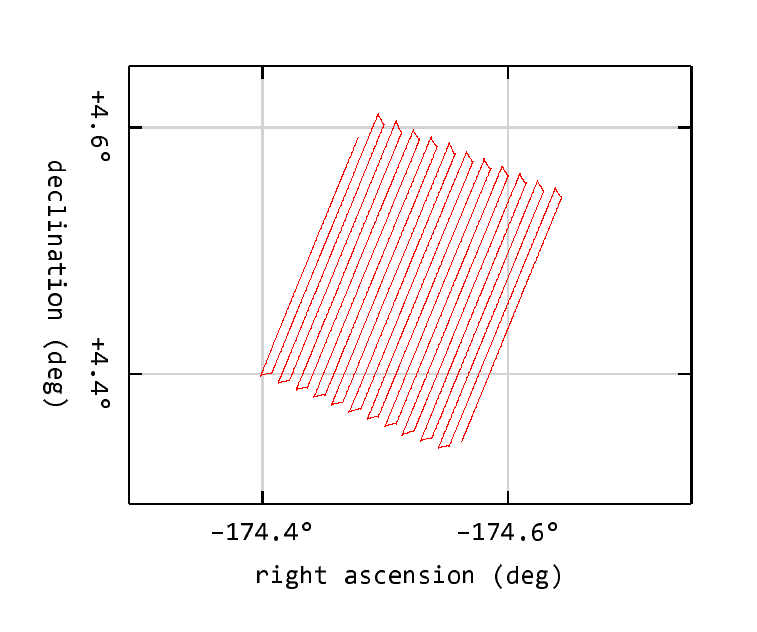}}\hfill
	\parbox[b]{0.35\linewidth}{\caption{Raw pointing data for a single PACS observation 1342225536. Telescope turn-arounds are visible at the ends of the straight scan legs. These turn-arounds are filtered out and the entire scan curve is split into individual scan legs.}}\hfill
\end{subfigure}

\vspace*{-0.5cm}

\begin{subfigure}[t]{\textwidth}
	\parbox{0.6\linewidth}{\includegraphics{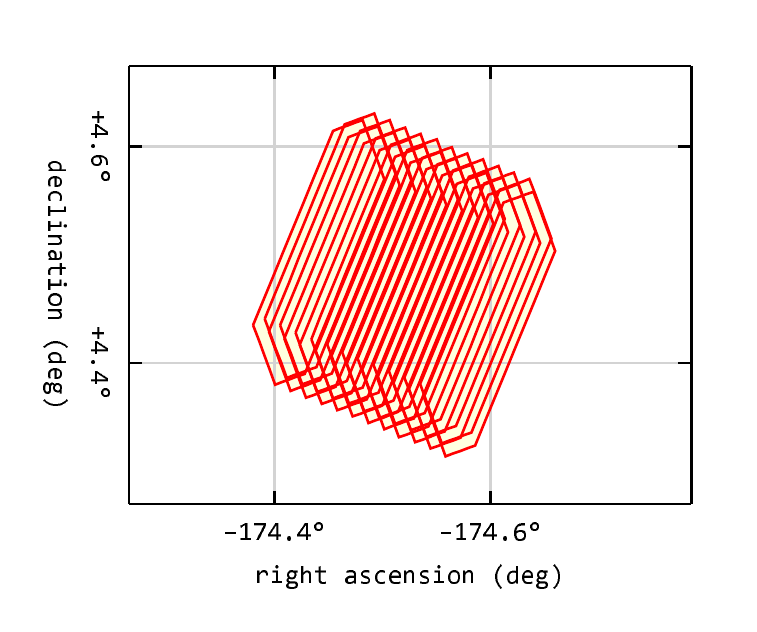}}\hfill
	\parbox[b]{0.35\linewidth}{\caption{Footprints of the individual scan legs after turn-around filtering. Scan leg footprints are determined by calculating the corners of the detector at the beginning and the end of each scan leg, \textit{cf.}~Panel~a), and taking the convex hull of the eight corner points.}}\hfill
\end{subfigure}

\vspace*{-0.5cm}

\begin{subfigure}[t]{\textwidth}
	\parbox{0.6\linewidth}{\includegraphics{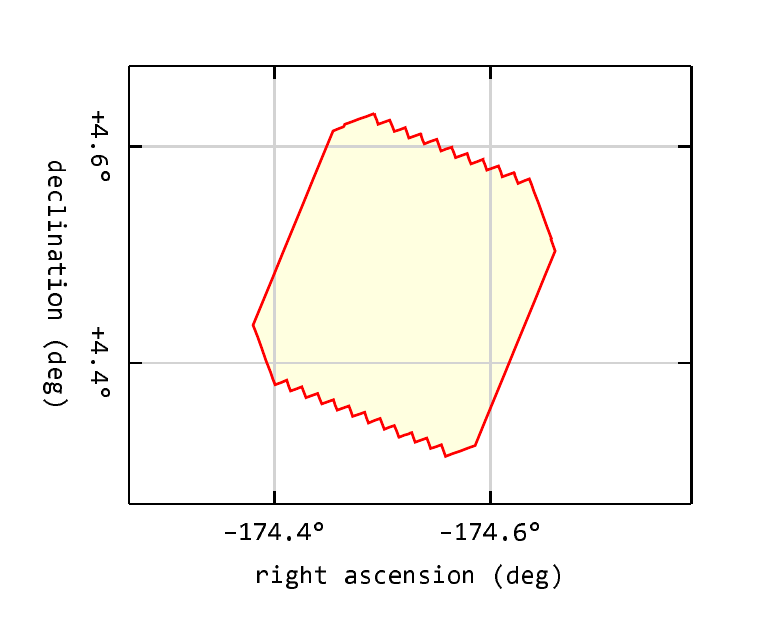}}\hfill
	\parbox[b]{0.35\linewidth}{\caption{Final footprint of the PACS scan map 1342225536, generated by taking the union of individual scan legs,\textit{cf.}~Panel~b). Calculating the exact union by removing overlapping regions is necessary to determine the area of the footprint.}}\hfill
\end{subfigure}

\vspace*{-0.25cm}
\caption{Reconstruction of the footprint of PACS scan map 1342225536 from raw pointing data. This particular scan map was taken in parallel mode with a repetition value of $1$. To achieve higher coverage, scan maps with perpendicular legs and multiple repetitions were also made.}
\label{fig:scanmap}
\end{figure}

We calculated the exact coverage of $26,761$ individual photometric observations made with the PACS and SPIRE instruments in scanning mode. No footprints were generated for failed or problematic observations (those which did not reach Level-1 processing). The scan maps were created directly from telescope pointing data. As an illustration, we plot a typical PACS scan map's raw pointings in panel a) of Fig.~\ref{fig:scanmap}. The straight lines of the trajectory (scan legs) are connected by the irregular sections where the telescope turned around to start the next leg.

After determining and filtering turn-arounds from the pointing data, using building block id (BbId)\footnote{The BbId is a number that identifies the building block (a consistent part) of an observation.} based selection we constructed the footprints of individual scan legs by taking the convex hull of the four corners of the detector at the beginning and at the end of each leg, see Panel~b) of Fig.~\ref{fig:scanmap}. The final footprint of an observation is obtained as the union of individual scan leg footprints, see Panel~c) of Fig.~\ref{fig:scanmap}. We made the assumption that the position angle of the detectors did not change during the scan with respect to the scan direction. In case of observations with multiple repetitions, thanks to the precise guidance of the spacecraft, scan legs of different repetition overlap almost exactly. For this reason, to simplify the computation of the union of scan legs, we generated the footprint from the very first repetition and omitted the rest. 

\begin{figure}

\vspace*{-0.3cm}

\begin{subfigure}[t]{\textwidth}
	\begin{center}
	\includegraphics{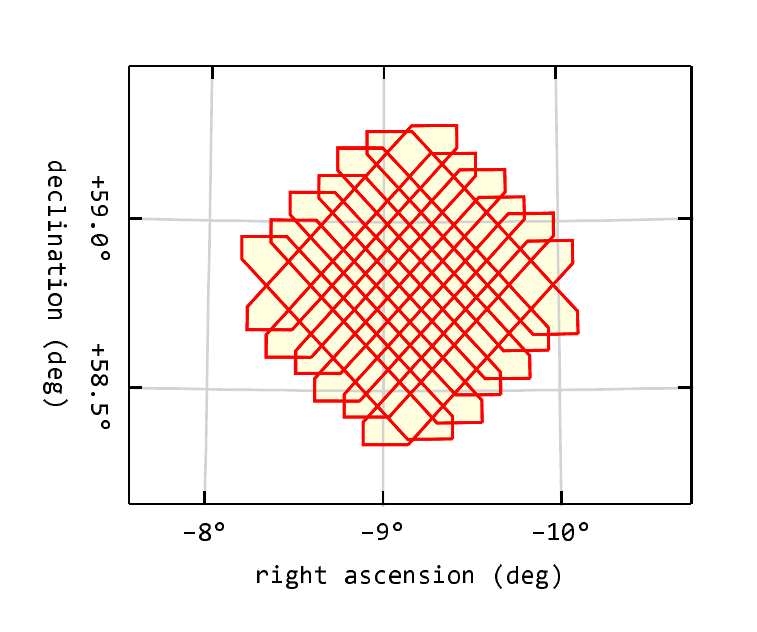}
	\end{center}
\end{subfigure}

\vspace*{-0.5cm}

\begin{subfigure}[t]{\textwidth}
	\begin{center}
	\includegraphics{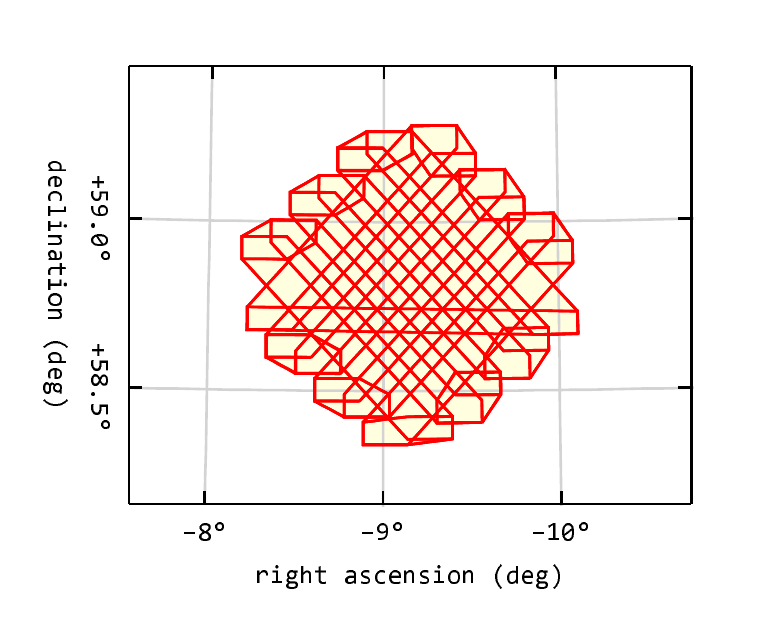}
	\end{center}
\end{subfigure}

	\caption{Scan legs of the SPIRE observation 1342183681 (top panel). As the scan map processing algorithm can make use of detector read outs during turn arounds, the final footprint is constructed by connecting the ends of successive scan legs (bottom panel).}
	\label{fig:spire_leg_ends}
\end{figure}

The SPIRE photometer was only used with observing modes where the telescope scanned along a sky region at a constant angular velocity: either 30 or $60 \arcsec/\unit{s}$. Maps are built by scanning at two predefined $\pm 42.4\degr$ ``magic'' angles, with respect to the instrument coordinate system. The different scans are also separated by a predefined offset. Nominally, the maps are built by scanning the two scan directions (i.e. cross-scans), providing redundancy and better coverage of the sky region. There are only a few cases of maps using a single scan direction only. For small map mode, there are only two shorts scans at an angle of $84.8\degr$, covering a sky area of $\sim 5\arcmin$ diameter. While in large map mode maps, the size of the sky region requested by the observer dictates the length of each scan leg and the number of scans in each scan direction. SPIRE scan map footprints were generated a way very similar to those of PACS with two exceptions. SPIRE pointing data was already filtered for turn around so we could skip that step. An important difference is that SPIRE scan maps are processed such a way that useful detector read outs during the turn arounds are also taken into account. For this reason, the corresponding ends of the successive scan legs are also connected, as it is illustrated in Fig.~\ref{fig:spire_leg_ends}.

In case of parallel mode, the PACS and SPIRE footprints were processed independently and the database contains the coverage of both instruments. In addition, we computed a parallel footprint which was taken to be the intersection of the two. In case of small, cross-scan maps the coverage is only homogeneous in the central part of the map and decreases significantly towards the edges. This information is available on a per pixel basis in final HSA data products but not included in the Footprint Database.

\subsection{Single point spectroscopy footprints}
\label{sec:spec_single}

\begin{figure}
	\begin{center}
	\includegraphics{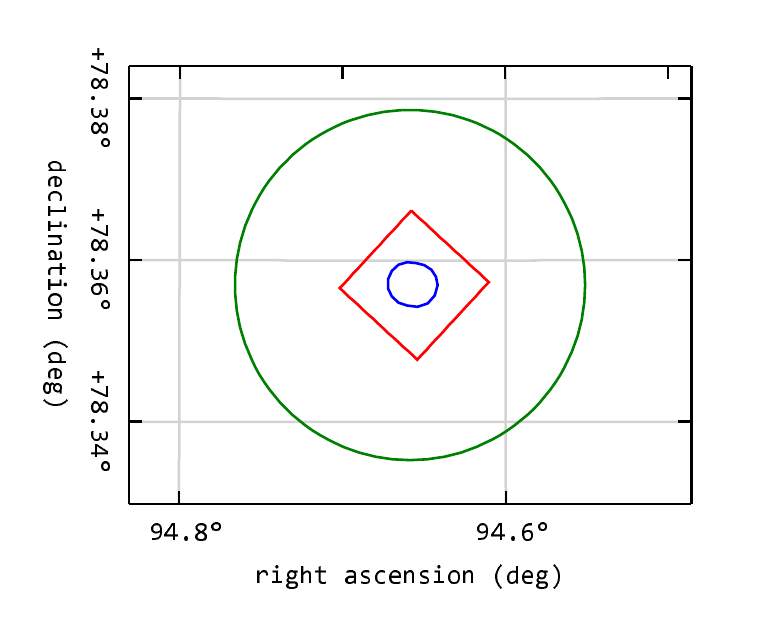}
	\end{center}
	\caption{Illustration of the different footprints of Herschel's spectroscopic instruments in single pointing mode. The combined field of view of the SPIRE beams is approximates by a circle with a diameter of $2\farcm 6$ (green). The footprint of the $5 \times 5$ spaxels of PACS, although organized in a slightly irregular pattern, is treated as a $47'' \times 47''$ square. HIFI beams are approximated by circles with a diameter equivalent to the half-power beam-width of the corresponding sub-bands. As HIFI beams of the two polarisations are slightly offset, the footprints are the union of two circles (blue).}
	\label{fig:spectro_pointed_all}
\end{figure}

\begin{figure}
	\begin{center}
	\includegraphics{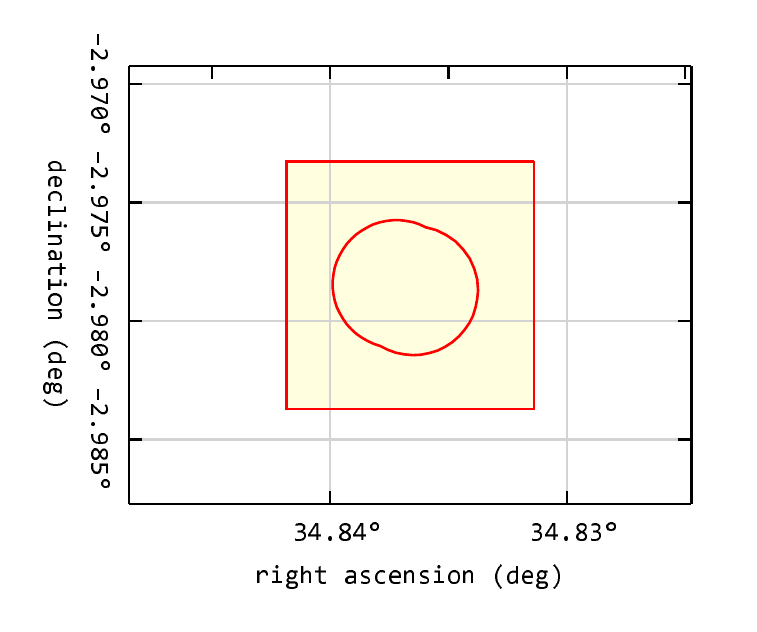}
	\end{center}
	\caption{Illustration of the footprints of a HIFI mapping (square region) and a pointed spectroscopic (oval) observation of Mira. The observation IDs are 1342262551 and 1342201114, respectively.}
	\label{fig:hifi_both}
\end{figure}

In Fig.~\ref{fig:spectro_pointed_all} we plot the typical footprints of the three spectroscopic instruments in single pointing mode. The footprint of the $5 \times 5$ spaxels of PACS is approximated by a $47'' \times 47''$ square, although one row of spaxels was not perfectly aligned with the rest. We used raw pointing data from Level-1 to identify calibration on and off positions but final footprint represent on positions only. The two SPIRE spectrometers have 19 and 37 horns arranged in a hexagonal pattern with an unvignetted field of view of $2'$ in diameter. Since SPIRE raster observations employed a step size of $116''$ along rows and $110''$ between the rows -- to avoid holes in the raster footprints, and to be consistent with raster observations -- we increased the representation of the field of view to $2\farcm 6$ in diameter. The footprints of single pointing HIFI observations are the union of two, slightly offset circles belonging to the two polarizations of the beam. The size of the offset and the half-power beam diameter depends on both, the polarization and the sub-band, cf.~Tab.~\ref{tab:spectroscopy}. The typical shape of HIFI single pointing footprints is better observable in Fig.~\ref{fig:hifi_both} where we plot it along with the footprint of a HIFI raster map. Footprints of HIFI spectral scan observations are always approximated by a single circle since -- at the time of building the footprint database -- no separate coordinates for the horizontal and vertical polarization positions were available in the HSA. This will be revised once the latest HSA products are available.

\subsection{Mapping spectroscopy footprints}
\label{sec:spec_mapping}

\begin{figure}

\vspace*{-0.3cm}

\begin{subfigure}[t]{\textwidth}
	\begin{center}
	\includegraphics{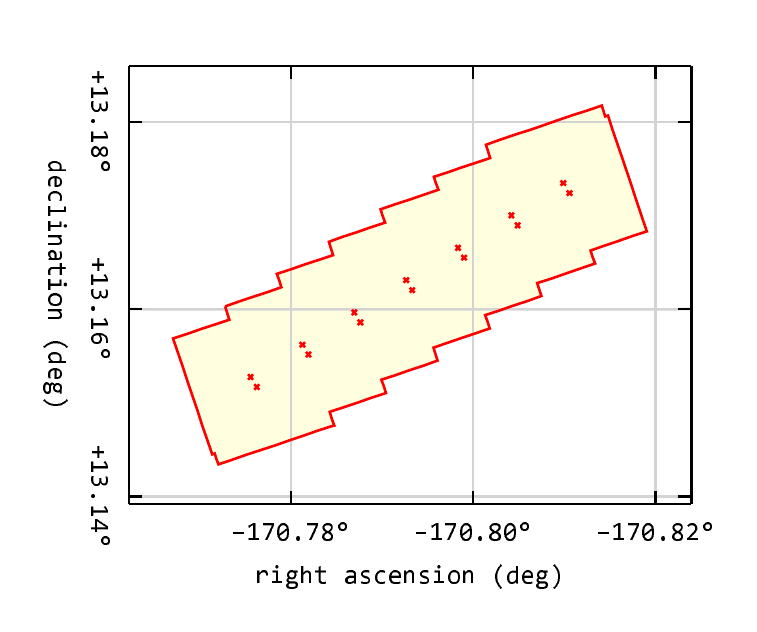}
	\end{center}
\end{subfigure}

\vspace*{-0.5cm}

\begin{subfigure}[t]{\textwidth}
	\begin{center}
	\includegraphics{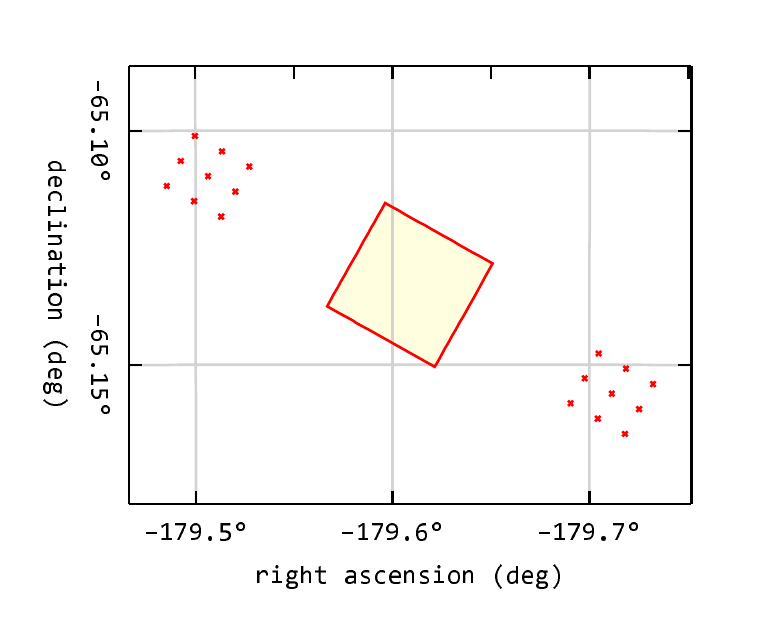}
	\end{center}
\end{subfigure}

	\caption{Typical PACS integral field spectroscopy raster maps. The filled area represents the observation footprint, the crosses show the telescope pointing. Observation 1342212598 (top panel) was made with no chopping while observation 1342240160 (bottom panel) used the chopper for differential measurement between the target and the reference points (crosses). Given that flags assigning to pointing products were erroneous, we had to implement heuristics to distinguish these types of observations from raw pointing information and average out coordinates appropriately to reconstruct the footprints.}
	\label{fig:pacs_raster}
\end{figure}

Reconstructing the footprint of PACS spectroscopy raster maps presented a challenge. Although detailed pointing data was available in HSA about the telescope movement during pointing to the raster points and the reference coordinates and during slewing between raster points, no meaningful flags were attached that would help organising pointing coordinates by raster row and column numbers. Also, most raster maps were measured with the chopper on but no information on the chopper throw was available. To identify the raster point coordinates and on and off positions, we turned to the friend of friend algorithm used for galaxy cluster finding. 
First, we grouped and averaged out pointings very close to each other to find raster points. Second, we implemented a heuristic algorithm to figure out whether the observation used the chopper or not and determined the observed positions accordingly. Fig.~\ref{fig:pacs_raster} illustrates two different types of PACS raster observations.

\begin{figure}
	\begin{center}
	\includegraphics{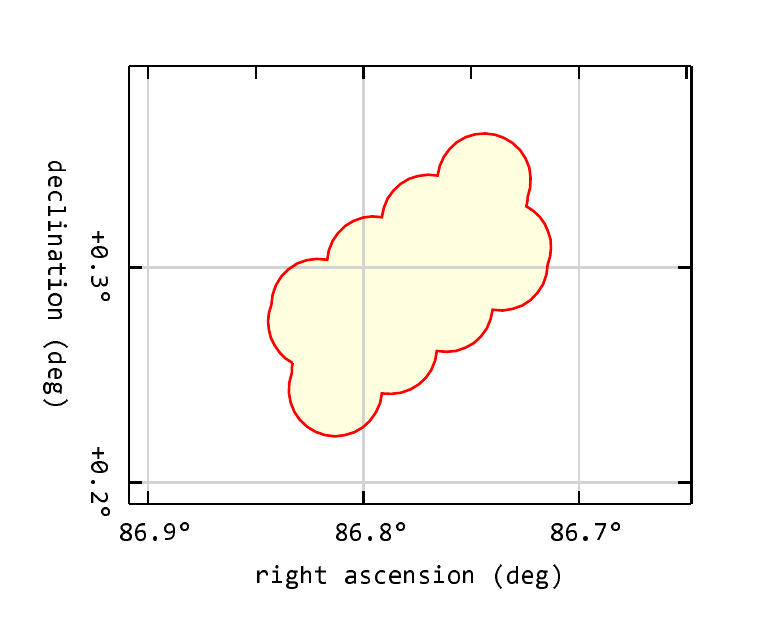}
	\end{center}
	\caption{Footprint of the SPIRE spectroscopic raster map observation 1342250523. The $2 \times 4$ raster map was repeated twice, with a small offset along the raster columns between the two repetitions.}
	\label{fig:spire_spectro_map}
\end{figure}

To construct the footprints of SPIRE spectroscopic raster maps, the list of raster points were extracted from Level-1 data using HIPE. Footprints are unions of circles of $2\farcm 6$ in diameter centred on the raster points, as illustrated in Fig.~\ref{fig:spire_spectro_map}. As it was mentioned in Sec.~\ref{sec:spec_single}, the $2\farcm 6$ diameter is larger than the unvignetted view of the detectors to avoid holes in the footprints.

The HIFI instrument was used to make spatially extended observations in two modes: on-the-fly maps and dual beam switch rasters. In each case, the size of the resulting rectangular maps are reported in the observation header. Since the reported map sizes correspond to the area scanned by the central chief ray of the PSF, we augmented the size of the rectangles by the radius of the half-power beam width in each direction. Recovering the exact orientation of HIFI maps presented a problem as the fly angle of the telescope was not correctly recorded during the early periods of the mission. Consequently, instead of using the rotation angles from the observation headers in HSA, we reached back to the database of original observation proposals, and used the angles as specified by the proposers. Fig.~\ref{fig:hifi_both} show the footprint of a typical HIFI map along with a pointed observation.

\subsection{Indexing the footprint database}

Since observation meta data is processed and organised into a uniform format, searching among all observations of all instruments based on header information is trivial. We provide a simple web-based interface to perform searches based on this header information. On the other hand, searching footprints based on spatial criteria is much more interesting. Spatial searches, in general, may include finding observations that cover a given pair of coordinates, observations that overlap with a given region, observations that cover a given region entirely, or observations that are covered by a given region completely, etc. 
As the exact determination of intersection or containment of complex spherical shapes is computationally intensive, comparing the query region with all scan map footprints is not possible every time a user request is processed. Instead, a smart indexing scheme is used to reduce computational resource requirements. Indexing is used primarily for coarse pre-filtering of matching candidates and exact intersection and containment is tested only for the good candidates. In case of a good indexing scheme, the reduction in computational time of spatial searches can be multiple orders of magnitude. We used the Spherical Library for SQL Server to build an index based on Hierarchical Triangular Mesh, described in details in \citet{Budavari2010}. In addition to simple point containment search, we implemented efficient intersection and region containment search for Herschel footprints. This is a novel result compared to existing footprint databases. For the details on the implementation, see Dobos~et~al. (2016, in preparation).

\subsection{Visualisation tools}

To visualise footprints, we developed a {\it new plotting library} for the spherical toolkit of \cite{Budavari2010}. The plotting library was written from scratch with flexibility and speed in mind. It supports numerous spherical projections (equirectangular, Aitoff, Hammer--Aitoff, Mollweide, orthographic, stereographic, etc.) and arbitrary rotation and zoom. The library was designed to provide publication quality output with default settings, for example, plots can be automatically rotated and zoomed such, that spherical shapes appear in the centre. Implementation of this feature is non-trivial and requires fast algorithms to estimate the minimal enclosing circle of the footprints and additional heuristics. For example, high speed plotting of projected arcs is achieved via adaptive refinement of poly-lines until they sufficiently approximate the exact curves. Multiple plotting layers can be used to display axis scales, borders, shapes, shape outlines, coordinate points with different symbols and grids. Axis scales can use either decimal or sexagesimal notation and tick and grid lines are also drawn accordingly. Figures of the present paper were generated with the new plotting library.

%

\section{Building the Herschel Footprint Database}
\label{sec:building}

\begin{figure}
	\center
	\includegraphics[clip=true,trim=0mm 130mm 184mm 0mm]{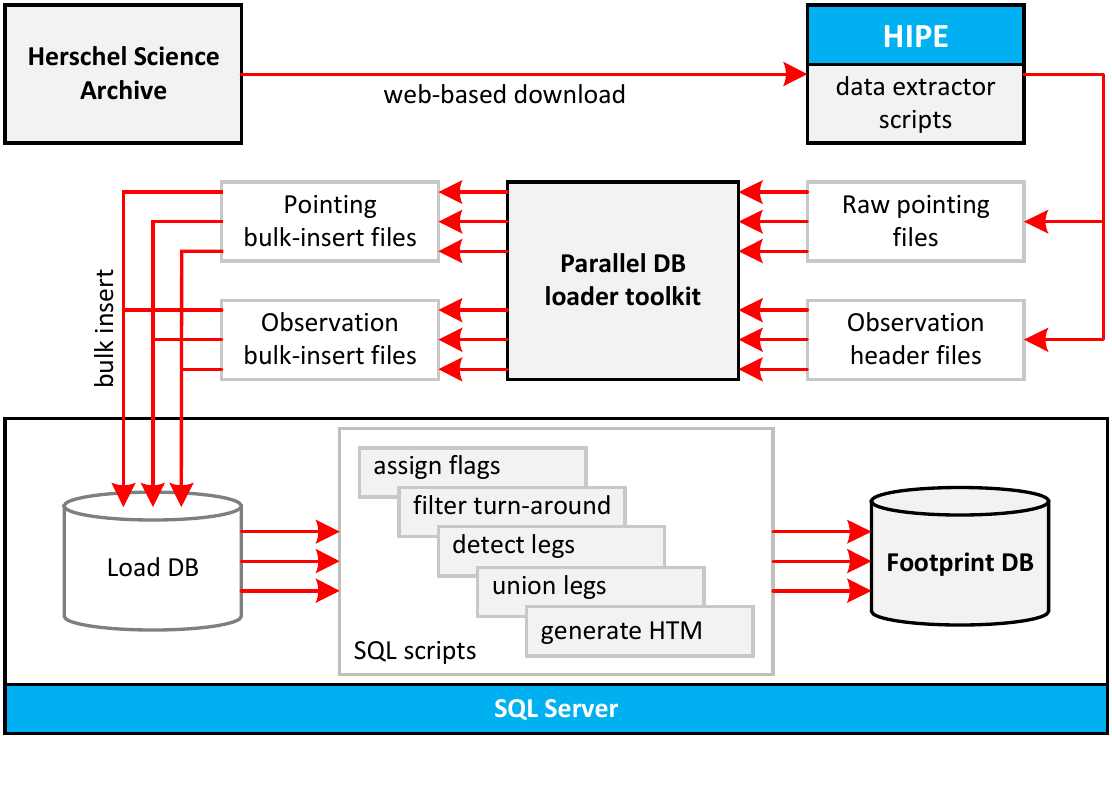}
	\caption{Flowchart of the data loading and footprint generation process. Data loading starts with downloading observation headers and processed Level~1 observations from the HSA using HIPE. Observation headers and pointing products are exported into text files for further processing by the DB loader toolkit. From this point all operations are parallelised to maximise system throughput. Raw output from HIPE is converted into binary files ready for database bulk-insert. Subsequent processing of pointing to generate footprints is entirely implemented in SQL. Multiple arrows mark paths of multi-threaded parallel data processing.}
	\label{fig:flowchart}
\end{figure}

Relational database management systems have been successfully used in many fields to store, search and process scientific data \citep{Szalay2002, Lemson2006, dobos2012}. We used Microsoft SQL Server~2012 to implement the footprint database. An excellent feature of this product is its powerful extensibility thanks to .Net framework integration. A significant part of the footprint processing logic and the entire Spherical Library was implemented as user-defined functions running inside the database. The web user interface and web services provide platform-independent access to the database. The size of the final database including indexes, is a modest 7.5~GB but an additional staging data space of 80~GB was required for processing raw pointing data. For data pre-processing, database ingestion and post-processing we implemented our own data loading pipeline. An overview of the data transformation procedure is drawn in Fig.~\ref{fig:flowchart}. 

To handle astronomical data, the efficient representation and indexing of spherical coordinates and shapes is of high importance. In Sec.~\ref{sec:regions}-\ref{sec:htm} we briefly review how this can be achieved. We also present novel solutions for reducing the complexity of spherical region descriptions in Sec.~\ref{sec:reduce}.

SQL scripts and source code are available on GitHub\footnote{\url{https://github.com/eltevo/HerschelDB}}.

\subsection{Data retrieval, processing and loading}
\label{sec:download}

Data from the Herschel space observatory is freely available for everyone but no uniformly processed catalogue of all observations has been published to date. Most of the Herschel data are automatically generated by systematic pipeline processing performed at the Herschel Science Centre (HSC) using the Herschel Data Processing system and made available through the HSA. A Herschel product may consist of observational products (scientific data), auxiliary products (spacecraft data), calibration products (parameters of the satellite and the instruments), quality control products (technical quality of the executed observation) and user provided data products (reduced data provided by the observers). The HSA is readily available from HIPE using Jython scripts.

We used Jython scripts in HIPE to retrieve the necessary observation headers and pointing information from the HSA and wrote them into ASCII files. The biggest challenge in data retrieval we faced was that the HIPE environment required up to 8~GB of RAM to open certain observations and process raw data to reconstruct pointing products. The typical time necessary to retrieve the pointing data for a scan map was as high as $2$ -- $3$~minutes. Further processing of the ASCII files was done by a parallel data processing framework developed by us to convert ASCII files into a format that can be ingested into a staging database using bulk-insert. Header information and pointing data from the various observational modes of the three instruments were converted into a common format at this step. 
The information about the schama of the resulting database available in our website, \url{http://herschel.vo.elte.hu}
The database loader toolkit was parallelised and executed on multiple threads to maximise the throughput of our multi-core system with high performance RAID storage. Once the data was loaded into the staging database, all subsequent data processing, except scan map generation, was done inside the database server by SQL scripts and user-defined functions.

\subsection{Representing spherical shapes}
\label{sec:regions}

To describe celestial regions in an analytic form, we use an upgraded version of the Spherical Library by \cite{Budavari2007, Budavari2010}. The software library was designed with astronomical use cases in mind, hence offers different functionality than typical geographic information systems (GIS) when it comes to spherical geometry. This includes double precision implementation for high accuracy and the ability to represent geometries not common in geography, e.g. a ring embracing the sphere along the equator. Fast calculation of area, boolean operations such as intersection and union, etc. are among the supported operations.

The Spherical Library represents the unit sphere in three dimensional Cartesian coordinates to circumvent problems arising from the singularities of the spherical coordinate system. Before going into the details of fast spatial search techniques, we briefly introduce the terminology of the software library which follows the three dimensional concept. By intersecting the unit sphere with a \textit{half space} one gets, in general, a small circle. If the half space crosses the centre of the sphere we get a great circle. By switching the sign of the half space the complementer of a small circle can be defined. More complex shapes can be described as intersections of multiple half spaces with the unit sphere. The intersection of half spaces is called a polyhedron which is always convex in the three dimensional space. On the other hand, the intersection of a polyhedron with the surface of the unit sphere is not necessarily convex in the spherical sense, i.e. not every two points of the shape can be connected by a great circle arc such a way that the shape contains the entire arc. Nevertheless, we will refer to the intersection of a polyhedron and the surface of the unit sphere as a \textit{convex}. To construct more complex shapes, the union of multiple convexes can be taken; we will refer to these as \textit{regions} and these will be roughly the equivalent of observation footprints. It is not necessary for a region to be connected, i.e. multiple, disjoint, non-adjacent patches are also a perfect region.

The Spherical Library is capable of \textit{simplifying} regions such a way that convexes making it up become disjoint. Once a region is completely simplified, the area and the list of bounding arcs, the so called \textit{outline} can be determined. A typical rectangular observation region is described with a single convex defined by four half spaces. In this case the half spaces go through the origin, thus the rectangle cut out from the sphere is bounded by great circles. In case of Herschel scan maps we obviously have much more complex regions.

\subsection{Hierarchical Triangular Mesh}
\label{sec:htm}

Relational database systems implement very efficient algorithms to sort and index data tables by integer keys. For the purpose of mapping coordinates and spherical shapes to integer number we use a technique called Hierarchical Triangular Mesh (HTM, \citealp{Kunszt2000}). We only briefly introduce HTM here, for more information refer to \citet{Budavari2010}.

Let us start with a regular octahedron inscribed into the unit sphere. By projecting the edges of the eight triangular sides of the octahedron, the spherical surface is divided into eight disjoint regions. Let us now split the triangles into four smaller ones by halving the edges and continue this a few dozen times. As the triangles get smaller the corresponding areas on the surface of the sphere will get smaller too. Twenty levels of recursion (or depth) is enough to reach the resolution of about $0\farcs3$. At the same time the algorithm to generate the triangles is extremely simple as it only requires halving sections in the three dimensional space and no trigonometric functions are needed to perform the calculations. The projections of triangles to the the spherical surface are called \textit{trixels}.

To map trixels to integer numbers a bit encoding scheme is used. A bit sequence identifying a specific trixel is called an \htmid{}. First we add a leading 1~bit to the very beginning to be able to tell the depth of the \htmid{} later. The eight sides of the initial octahedron require three bits to encode which become the next three most significant bits of the \htmid{}. Every additional level of recursion requires two additional bits that are written after the existing ones. The recursively generated trixels define a tree structure: every larger trixel is split into four smaller ones which is called a quad-tree. Encoding the tree into a bit sequence as we did it is called Morton coding. Strictly speaking, \htmid{}s refer to small areas on the sphere instead of exact coordinates and it is possible for different coordinates to have the same corresponding \htmid{} if the recursion level is low. In typical astronomical applications 44 bits (20 levels) are far enough to encode spherical coordinate pairs. Database systems provide the 64~bit \texttt{bigint} data-type to store such numbers, and although it seems a waste of bits, technology exists to compress leading zeros and save on disk space. 

\begin{figure}
	\center
	\includegraphics{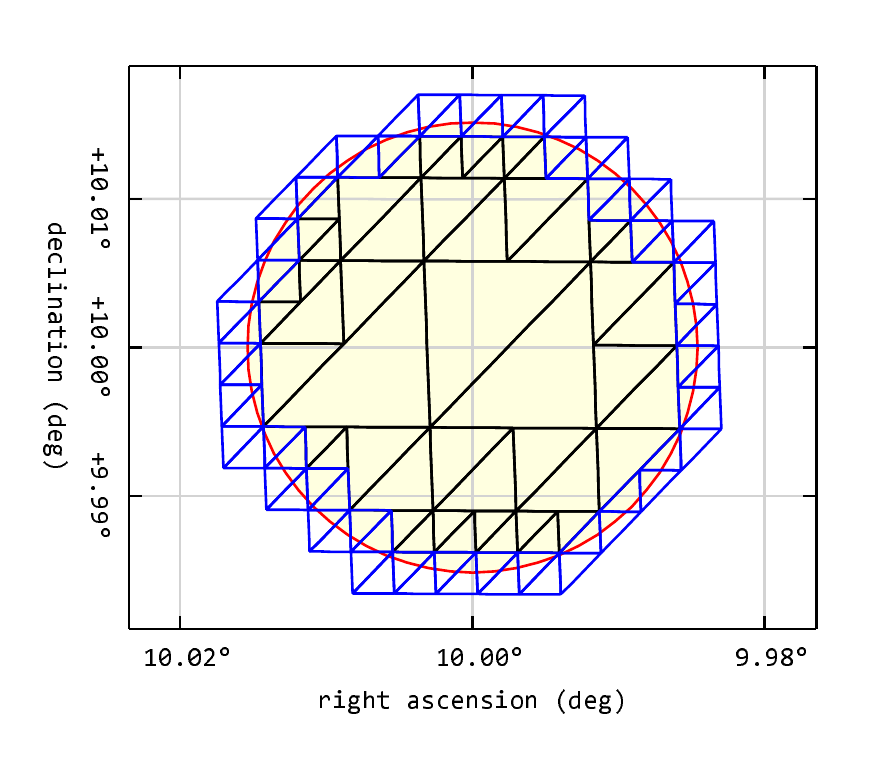}
	\caption{Hierarchical triangular mesh covering of a sample region. The original circle is in red, fully covered trixels are black and partial trixels lying on the boundary are blue. Note, that while fully covered trixels can be larger (i.e. they are on an upper level of the HTM tree), partial trixels are always the same size. As boundaries of regions seldom coincide with trixel edges, a lower limit on trixel size is necessary to control the number of \htmid{} ranges covering a region.}
	\label{fig:htmcover}
\end{figure}

We have already shown how to hash coordinate pairs into the \htmid{} but to support searches that involve spherical shapes we need one more step which is calculating the \textit{cover} of shapes. As it was discussed in Sec.~\ref{sec:regions}, spherical shapes (or regions) are parts of the spherical surface bounded by closed arcs of small or great circles. We are now looking for a minimal set of trixels that cover a particular region. First we test which top-level trixels of the eight are necessary to cover the entire region and then we recursively refine to approximate the shape with trixels as much as possible, but at the same time, we want to keep the number of \htmid{} intervals as low as possible. During a refinement test, corners of the trixels are tested whether they are inside or outside the region. This can be done quite quickly as regions are defined by half-spaces, so the only computation involved is solving linear inequalities. Once a trixel is found to be completely inside the region it is added to the list of full trixels. Shall a trixel fall on the boundary of the shape, it is split into four smaller trixels which are tested against the region one by one and split further if necessary. Trixels falling entirely outside the region are discarded. Fig.~\ref{fig:htmcover} shows a typical HTM cover of a spherical shape. The final result of a cover, once trixels are converted into \htmid{} ranges, is a set of disjoint intervals of positive numbers. After some merging and pruning, the intervals can be used to perform spatial search using traditional SQL queries.

\subsection{Reducing the complexity of scan map footprints}
\label{sec:reduce}

Footprints of large Herschel scan maps tend to be quite complicated when described as a spherical region, c.f. Sec.~\ref{sec:regions}, due to the jagged edges. Therefore, we implemented two simple methods to reduce the complexity of such regions for applications where computational efficiency in more important than precision. Both methods start with the outline of the regions, i.e. a continuous loop composed of arcs of great circles. (We note that the Spherical Library also supports using arcs of small circles. The simplification methods described below cannot directly be used for such regions but there are ways of generalisation.) The first method works by straightening small kinks of the outline by the Douglas--Peucker algorithm, the second method eliminates jagged scan map edges by taking the convex hull of the region. The former is applicable when approximate visualisation of the footprints is satisfactory, whereas the convex hull of the footprints can be used for searches when false positive matches are acceptable.

\begin{figure}

\vspace*{-0.3cm}

\begin{subfigure}[b]{\textwidth}
	\parbox{0.6\linewidth}{\includegraphics{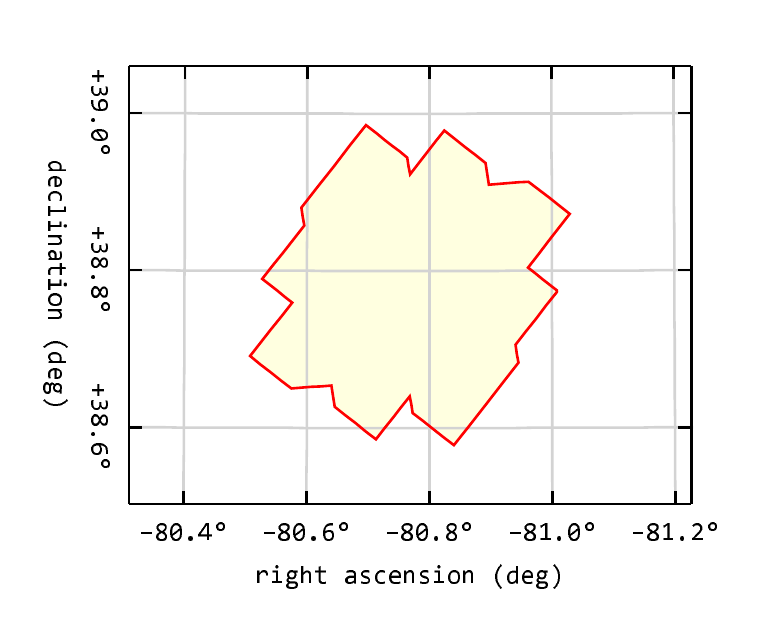}}\hfill
	\label{fig:orig}
	\parbox[b]{0.35\linewidth}{\caption{The original footprint without simplification.}}\hfill
\end{subfigure}

\vspace*{-0.5cm}

\begin{subfigure}[b]{\textwidth}
	\parbox{0.6\linewidth}{\includegraphics{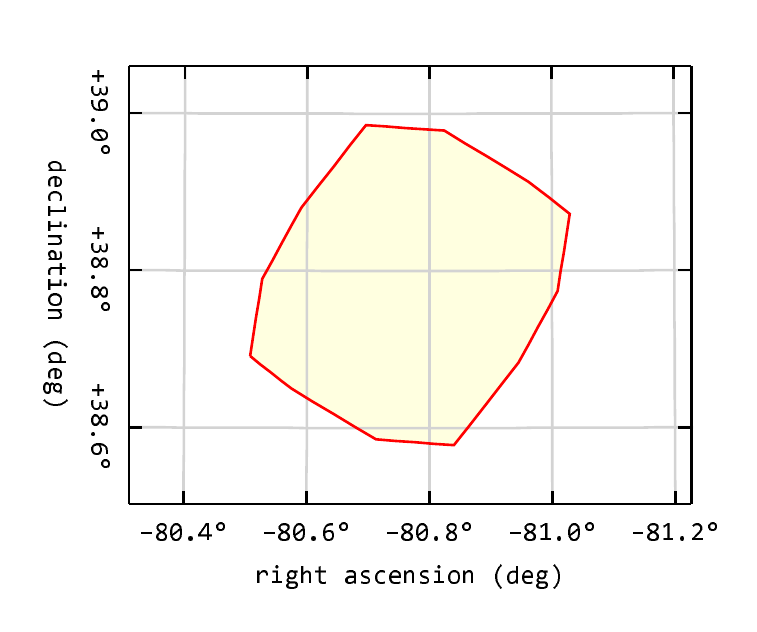}}\hfill
	\label{fig:chull}
	\parbox[b]{0.35\linewidth}{\caption{Convex hull of the footprint. The convex hull can be relatively easily calculate for any footprint bounded by great circles and smaller than the hemisphere. The area of the convex hull is always an upper approximation of the original area.}}\hfill
\end{subfigure}

\vspace*{-0.5cm}

\begin{subfigure}[b]{\textwidth}
	\parbox{0.6\linewidth}{\includegraphics{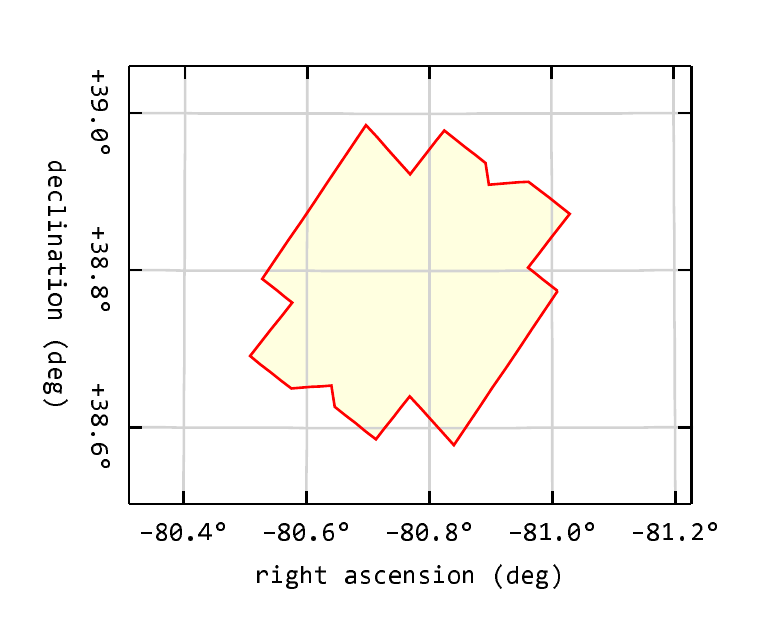}}\hfill
	\label{fig:dougpeuck}
	\parbox[b]{0.35\linewidth}{\caption{The Douglas--Peucker reduction of the outline. The limit $\epsilon$ was set to $50$~arc~sec to emphasize the effect.}}\hfill
\end{subfigure}

\vspace*{-0.25cm}

\caption{Footprints of the SPIRE observation 1342186861 using two different reduction methods.}
\label{fig:simplify}
\end{figure}

\subsubsection{The Douglas--Peucker reduction of the outlines.} 

The Douglas--Peucker algorithm \citep{DouglasPeucker1973} was originally created for reducing the complexity of a planar curve composed of straight line segments. We generalised the algorithm for spherical curves consisting of great circle arcs the following way. The procedure starts by taking all arc endpoints and orders them according to their positions along the curve. If the curve is a closed loop, the starting and ending points will be the same and are usually chosen arbitrarily, thus the reduction of a curve might not always be the same. The algorithm works recursively by considering the start and end points first. A recursive steps consists of finding between the starting $A$ and ending $B$ points the point $C$ that lies the furthest (in terms of spherical distance) from the great circle arc $\nicearc{AB}$ connecting $A$ and $B$ directly. If $C$ is closer to $\nicearc{AB}$ than a predefined angle $\epsilon$, all points between $A$ and $B$ are discarded and the entire section is substituted with $\nicearc{AB}$ and the algorithm stops. If $C$ is further from $\nicearc{AB}$ than $\epsilon$ then the algorithm is called recursively for $\nicearc{AC}$ and $\nicearc{CB}$. Panel~c) of Fig.~\ref{fig:simplify} illustrates the results of the spherical Douglas--Peucker algorithm performed on a SPIRE scan map footprint with $\epsilon = 50''$. As it is clearly visible, the small wiggles of the outline are eliminated and substituted with single arcs. While the reduced outline looks certainly simpler, it is usually just a coarse approximation of the original and estimating the area of the original footprint based on the reduction is problematic.

\subsubsection{The convex hull method.} 

To determine the convex hull of a spherical region bounded by arcs of great circles, we start by determining the arc endpoints. The convex hull of a set of points on the surface of the square is well defined if the points are confined to a hemisphere only. Instead of solving the convex hull problem directly on the spherical surface, we solve the following, equivalent problem. We take the Cartesian unit vectors pointing to the arc endpoints plus the origin of the sphere and compute the 3-dimensional convex hull of these using the QuickHull algorithm \citep{Quickhull}. By enumerating the faces of the 3D convex hull, one can find the great circle arcs bounding the spherical convex hull. Panel~b) of Fig.~\ref{fig:simplify} illustrates the convex hull of a SPIRE scan map calculated from the original scan map plotted in Panel~a) of Fig.~\ref{fig:simplify}. The convex hull method reduces complexity by ``filling in'' concave parts of the footprint which is not appropriate for small cross scans, for example. The area of the convex hull is always an upper estimate of the area of the original footprint.

\subsubsection{Accessing the database form scripts.}

Scripting access to the database is possible via the REST web service we provide at \url{http://herschel.vo.elte.hu}. The service can return the data in plain text, JSON and XML formats. The Python script in Listing~\ref{python:plot}\footnote{The script was written and tested on Python 3.5.1. For further information see the official home of the Python Programming Language at \url{https://www.python.org/}} shows how to query the observation list and plot the footprints. The complete documentation of the RESTful web service API, as well as the description of the database schema, is available on the web page.

\vspace*{0.5cm}

\begin{python}
\begin{lstlisting}
import matplotlib.pyplot as plt
import numpy as np
import urllib.parse
import urllib.request
import json

baseURL = "http://herschel.vo.elte.hu/Api/"

def openURL(url, values, headers):
    if values == None:
        url = baseURL + url
    else:
        params = urllib.parse.urlencode(values)
        url = baseURL + url + '?' + params
    req = urllib.request.Request(url, headers=headers)
    with urllib.request.urlopen(req) as response:
        data = response.read()
        data = data.decode("ascii")
        return data
    
def openTable(url, params):
    data = openURL(url, params, {'Accept': 'text/plain'})
    return np.loadtxt(data.splitlines())

def openJSON(url, params):
    data = openURL(url, params, {'Accept': 'application/json'})
    return json.loads(data)

def findObsIntersect(inst, region):
    url = "Observations"
    params = { "findby": "intersect", "inst": inst, 
               "region": region }
    return openJSON(url, params)

def getOutlinePoints(inst, obsID, res):
    url = "Observations/" + inst + "/" + str(obsID) + 
    	  "/Footprint/Outline/Points"
    params = { "res": res }
    return openTable(url, params)
    
obs = findObsIntersect("PACS", "CIRCLE J2000 207.25 -28.4 20")
for o in obs:
    points = getOutlinePoints(o["inst"], o["obsID"], 0.2)
    plt.plot(points[:,1], points[:,2], '-')
plt.show()
\end{lstlisting}
\caption{ simple python script to demonstrate how easy it is to retrieve observation footprints intersecting with a circle of a radius of $20\arcmin$, centred on $(\alpha, \delta) = 207\fdg 25, -28 \fdg 4 $. Once a list of observations is retrieved by the function \texttt{findObsPoints}, points of the footprints are also queried using \texttt{getOutlinePoints}. The parameter \texttt{res} determines the resampling resolution of small circle arcs to great circle arcs. The complete documentation of the RESTful web service API is available on the web page.}
\label{python:plot}
\end{python}

\section{Summary}
\label{sec:summary}

We have built a database containing meta data and footprint of all observations made by the Herschel Space Observatory during its entire mission between 2009 and 2013. Meta-data of the many observational modes of the three instruments have been homogenized to fit into a single relational data model. By using telescope pointing information and other meta-data, we have reconstructed of the footprint of observations including single pointing, raster maps and scan maps. The footprint are represented as non-pixellated, exact description of the observed areas. Since scan map footprint can become very complicated for many application, we have implemented two simplification algorithms to reduce the footprint outlines. The database is accessible via a web page and a REST web service at \url{http://herschel.vo.elte.hu}


\bibliographystyle{spbasic}      
\bibliography{herschel-expa}

\begin{thebibliography}{20}
\providecommand{\natexlab}[1]{#1}
\providecommand{\url}[1]{{#1}}
\providecommand{\urlprefix}{URL }
\expandafter\ifx\csname urlstyle\endcsname\relax
  \providecommand{\doi}[1]{DOI~\discretionary{}{}{}#1}\else
  \providecommand{\doi}{DOI~\discretionary{}{}{}\begingroup
  \urlstyle{rm}\Url}\fi
\providecommand{\eprint}[2][]{\url{#2}}

\bibitem[{{Balog} et~al(2014){Balog}, {M{\"u}ller}, {Nielbock}, {Altieri},
  {Klaas}, {Blommaert}, {Linz}, {Lutz}, {Mo{\'o}r}, {Billot}, {Sauvage}, and
  {Okumura}}]{Balog2014}
{Balog} Z, {M{\"u}ller} T, {Nielbock} M, {Altieri} B, {Klaas} U, {Blommaert} J,
  {Linz} H, {Lutz} D, {Mo{\'o}r} A, {Billot} N, {Sauvage} M, {Okumura} K (2014)
  {The Herschel-PACS photometer calibration. Point-source flux calibration for
  scan maps}. Experimental Astronomy 37:129--160,
  \doi{10.1007/s10686-013-9352-3}, \eprint{1309.6099}

\bibitem[{Barber et~al(1996)Barber, Dobkin, and Huhdanpaa}]{Quickhull}
Barber CB, Dobkin DP, Huhdanpaa H (1996) The quickhull algorithm for convex
  hulls. ACM Transactions on Mathematical Software (TOMS) 22(4):469--483

\bibitem[{{Budav{\'a}ri} et~al(2007){Budav{\'a}ri}, {Dobos}, {Szalay},
  {Greene}, {Gray}, and {Rots}}]{Budavari2007}
{Budav{\'a}ri} T, {Dobos} L, {Szalay} AS, {Greene} G, {Gray} J, {Rots} AH
  (2007) {Footprint Services for Everyone}. In: {Shaw} RA, {Hill} F, {Bell} DJ
  (eds) Astronomical Data Analysis Software and Systems XVI, Astronomical
  Society of the Pacific Conference Series, vol 376, p 559

\bibitem[{{Budav{\'a}ri} et~al(2010){Budav{\'a}ri}, {Szalay}, and
  {Fekete}}]{Budavari2010}
{Budav{\'a}ri} T, {Szalay} AS, {Fekete} G (2010) {Searchable Sky Coverage of
  Astronomical Observations: Footprints and Exposures}. \pasp 122:1375--1388,
  \doi{10.1086/657302}, \eprint{1005.2606}

\bibitem[{Budavari et~al(2013)Budavari, Dobos, and Szalay}]{SkyQuery}
Budavari T, Dobos L, Szalay AS (2013) Skyquery: Federating astronomy archives.
  Computing in Science and Engineering 15(3):12--20,
  \doi{10.1109/MCSE.2013.41},
  \urlprefix\url{http://doi.ieeecomputersociety.org/10.1109/MCSE.2013.41}

\bibitem[{{de Graauw} et~al(2010){de Graauw}, {Helmich}, {Phillips}, {Stutzki},
  {Caux}, {Whyborn}, {Dieleman}, {Roelfsema}, {Aarts}, {Assendorp},
  {Bachiller}, {Baechtold}, {Barcia}, {Beintema}, {Belitsky}, {Benz}, {Bieber},
  {Boogert}, {Borys}, {Bumble}, {Ca{\"i}s}, {Caris}, {Cerulli-Irelli},
  {Chattopadhyay}, {Cherednichenko}, {Ciechanowicz}, {Coeur-Joly}, {Comito},
  {Cros}, {de Jonge}, {de Lange}, {Delforges}, {Delorme}, {den Boggende},
  {Desbat}, {Diez-Gonz{\'a}lez}, {di Giorgio}, {Dubbeldam}, {Edwards},
  {Eggens}, {Erickson}, {Evers}, {Fich}, {Finn}, {Franke}, {Gaier}, {Gal},
  {Gao}, {Gallego}, {Gauffre}, {Gill}, {Glenz}, {Golstein}, {Goulooze},
  {Gunsing}, {G{\"u}sten}, {Hartogh}, {Hatch}, {Higgins}, {Honingh}, {Huisman},
  {Jackson}, {Jacobs}, {Jacobs}, {Jarchow}, {Javadi}, {Jellema}, {Justen},
  {Karpov}, {Kasemann}, {Kawamura}, {Keizer}, {Kester}, {Klapwijk}, {Klein},
  {Kollberg}, {Kooi}, {Kooiman}, {Kopf}, {Krause}, {Krieg}, {Kramer},
  {Kruizenga}, {Kuhn}, {Laauwen}, {Lai}, {Larsson}, {Leduc}, {Leinz}, {Lin},
  {Liseau}, {Liu}, {Loose}, {L{\'o}pez-Fernandez}, {Lord}, {Luinge}, {Marston},
  {Mart{\'{\i}}n-Pintado}, {Maestrini}, {Maiwald}, {McCoey}, {Mehdi}, {Megej},
  {Melchior}, {Meinsma}, {Merkel}, {Michalska}, {Monstein}, {Moratschke},
  {Morris}, {Muller}, {Murphy}, {Naber}, {Natale}, {Nowosielski}, {Nuzzolo},
  {Olberg}, {Olbrich}, {Orfei}, {Orleanski}, {Ossenkopf}, {Peacock}, {Pearson},
  {Peron}, {Phillip-May}, {Piazzo}, {Planesas}, {Rataj}, {Ravera}, {Risacher},
  {Salez}, {Samoska}, {Saraceno}, {Schieder}, {Schlecht}, {Schl{\"o}der},
  {Schm{\"u}lling}, {Schultz}, {Schuster}, {Siebertz}, {Smit}, {Szczerba},
  {Shipman}, {Steinmetz}, {Stern}, {Stokroos}, {Teipen}, {Teyssier}, {Tils},
  {Trappe}, {van Baaren}, {van Leeuwen}, {van de Stadt}, {Visser}, {Wildeman},
  {Wafelbakker}, {Ward}, {Wesselius}, {Wild}, {Wulff}, {Wunsch}, {Tielens},
  {Zaal}, {Zirath}, {Zmuidzinas}, and {Zwart}}]{HIFI}
{de Graauw} T, {Helmich} FP, {Phillips} TG, {Stutzki} J, {Caux} E, {Whyborn}
  ND, {Dieleman} P, {Roelfsema} PR, {Aarts} H, {Assendorp} R, {Bachiller} R,
  {Baechtold} W, {Barcia} A, {Beintema} DA, {Belitsky} V, {Benz} AO, {Bieber}
  R, {Boogert} A, {Borys} C, {Bumble} B, {Ca{\"i}s} P, {Caris} M,
  {Cerulli-Irelli} P, {Chattopadhyay} G, {Cherednichenko} S, {Ciechanowicz} M,
  {Coeur-Joly} O, {Comito} C, {Cros} A, {de Jonge} A, {de Lange} G, {Delforges}
  B, {Delorme} Y, {den Boggende} T, {Desbat} JM, {Diez-Gonz{\'a}lez} C, {di
  Giorgio} AM, {Dubbeldam} L, {Edwards} K, {Eggens} M, {Erickson} N, {Evers} J,
  {Fich} M, {Finn} T, {Franke} B, {Gaier} T, {Gal} C, {Gao} JR, {Gallego} JD,
  {Gauffre} S, {Gill} JJ, {Glenz} S, {Golstein} H, {Goulooze} H, {Gunsing} T,
  {G{\"u}sten} R, {Hartogh} P, {Hatch} WA, {Higgins} R, {Honingh} EC, {Huisman}
  R, {Jackson} BD, {Jacobs} H, {Jacobs} K, {Jarchow} C, {Javadi} H, {Jellema}
  W, {Justen} M, {Karpov} A, {Kasemann} C, {Kawamura} J, {Keizer} G, {Kester}
  D, {Klapwijk} TM, {Klein} T, {Kollberg} E, {Kooi} J, {Kooiman} PP, {Kopf} B,
  {Krause} M, {Krieg} JM, {Kramer} C, {Kruizenga} B, {Kuhn} T, {Laauwen} W,
  {Lai} R, {Larsson} B, {Leduc} HG, {Leinz} C, {Lin} RH, {Liseau} R, {Liu} GS,
  {Loose} A, {L{\'o}pez-Fernandez} I, {Lord} S, {Luinge} W, {Marston} A,
  {Mart{\'{\i}}n-Pintado} J, {Maestrini} A, {Maiwald} FW, {McCoey} C, {Mehdi}
  I, {Megej} A, {Melchior} M, {Meinsma} L, {Merkel} H, {Michalska} M,
  {Monstein} C, {Moratschke} D, {Morris} P, {Muller} H, {Murphy} JA, {Naber} A,
  {Natale} E, {Nowosielski} W, {Nuzzolo} F, {Olberg} M, {Olbrich} M, {Orfei} R,
  {Orleanski} P, {Ossenkopf} V, {Peacock} T, {Pearson} JC, {Peron} I,
  {Phillip-May} S, {Piazzo} L, {Planesas} P, {Rataj} M, {Ravera} L, {Risacher}
  C, {Salez} M, {Samoska} LA, {Saraceno} P, {Schieder} R, {Schlecht} E,
  {Schl{\"o}der} F, {Schm{\"u}lling} F, {Schultz} M, {Schuster} K, {Siebertz}
  O, {Smit} H, {Szczerba} R, {Shipman} R, {Steinmetz} E, {Stern} JA, {Stokroos}
  M, {Teipen} R, {Teyssier} D, {Tils} T, {Trappe} N, {van Baaren} C, {van
  Leeuwen} BJ, {van de Stadt} H, {Visser} H, {Wildeman} KJ, {Wafelbakker} CK,
  {Ward} JS, {Wesselius} P, {Wild} W, {Wulff} S, {Wunsch} HJ, {Tielens} X,
  {Zaal} P, {Zirath} H, {Zmuidzinas} J, {Zwart} F (2010) {The
  Herschel-Heterodyne Instrument for the Far-Infrared (HIFI)}. \aap 518:L6,
  \doi{10.1051/0004-6361/201014698}

\bibitem[{{Dobos} et~al(2004){Dobos}, {Budav{\'a}ri}, {Csabai}, and
  {Szalay}}]{Dobos2004}
{Dobos} L, {Budav{\'a}ri} T, {Csabai} I, {Szalay} AS (2004) {Spectrum and
  Bandpass Services for the Virtual Observatory}. In: {Ochsenbein} F, {Allen}
  MG, {Egret} D (eds) Astronomical Data Analysis Software and Systems (ADASS)
  XIII, Astronomical Society of the Pacific Conference Series, vol 314, p 185

\bibitem[{Dobos et~al(2012)Dobos, Budav{\'a}ri, Li, Szalay, and
  Csabai}]{dobos2012}
Dobos L, Budav{\'a}ri T, Li N, Szalay AS, Csabai I (2012) Skyquery: an
  implementation of a parallel probabilistic join engine for
  cross-identification of multiple astronomical databases. In: Scientific and
  Statistical Database Management, Springer, pp 159--167

\bibitem[{Douglas and Peucker(1973)}]{DouglasPeucker1973}
Douglas DH, Peucker TK (1973) Algorithms for the reduction of the number of
  points required to represent a digitized line or its caricature.
  Cartographica: The International Journal for Geographic Information and
  Geovisualization 10(2):112--122

\bibitem[{{Griffin} et~al(2010){Griffin}, {Abergel}, {Abreu}, {Ade},
  {Andr{\'e}}, {Augueres}, {Babbedge}, {Bae}, {Baillie}, {Baluteau}, {Barlow},
  {Bendo}, {Benielli}, {Bock}, {Bonhomme}, {Brisbin}, {Brockley-Blatt},
  {Caldwell}, {Cara}, {Castro-Rodriguez}, {Cerulli}, {Chanial}, {Chen},
  {Clark}, {Clements}, {Clerc}, {Coker}, {Communal}, {Conversi}, {Cox},
  {Crumb}, {Cunningham}, {Daly}, {Davis}, {de Antoni}, {Delderfield}, {Devin},
  {di Giorgio}, {Didschuns}, {Dohlen}, {Donati}, {Dowell}, {Dowell}, {Duband},
  {Dumaye}, {Emery}, {Ferlet}, {Ferrand}, {Fontignie}, {Fox}, {Franceschini},
  {Frerking}, {Fulton}, {Garcia}, {Gastaud}, {Gear}, {Glenn}, {Goizel},
  {Griffin}, {Grundy}, {Guest}, {Guillemet}, {Hargrave}, {Harwit}, {Hastings},
  {Hatziminaoglou}, {Herman}, {Hinde}, {Hristov}, {Huang}, {Imhof}, {Isaak},
  {Israelsson}, {Ivison}, {Jennings}, {Kiernan}, {King}, {Lange}, {Latter},
  {Laurent}, {Laurent}, {Leeks}, {Lellouch}, {Levenson}, {Li}, {Li},
  {Lilienthal}, {Lim}, {Liu}, {Lu}, {Madden}, {Mainetti}, {Marliani}, {McKay},
  {Mercier}, {Molinari}, {Morris}, {Moseley}, {Mulder}, {Mur}, {Naylor},
  {Nguyen}, {O'Halloran}, {Oliver}, {Olofsson}, {Olofsson}, {Orfei}, {Page},
  {Pain}, {Panuzzo}, {Papageorgiou}, {Parks}, {Parr-Burman}, {Pearce},
  {Pearson}, {P{\'e}rez-Fournon}, {Pinsard}, {Pisano}, {Podosek}, {Pohlen},
  {Polehampton}, {Pouliquen}, {Rigopoulou}, {Rizzo}, {Roseboom}, {Roussel},
  {Rowan-Robinson}, {Rownd}, {Saraceno}, {Sauvage}, {Savage}, {Savini},
  {Sawyer}, {Scharmberg}, {Schmitt}, {Schneider}, {Schulz}, {Schwartz},
  {Shafer}, {Shupe}, {Sibthorpe}, {Sidher}, {Smith}, {Smith}, {Smith},
  {Spencer}, {Stobie}, {Sudiwala}, {Sukhatme}, {Surace}, {Stevens}, {Swinyard},
  {Trichas}, {Tourette}, {Triou}, {Tseng}, {Tucker}, {Turner}, {Vaccari},
  {Valtchanov}, {Vigroux}, {Virique}, {Voellmer}, {Walker}, {Ward}, {Waskett},
  {Weilert}, {Wesson}, {White}, {Whitehouse}, {Wilson}, {Winter}, {Woodcraft},
  {Wright}, {Xu}, {Zavagno}, {Zemcov}, {Zhang}, and {Zonca}}]{SPIRE}
{Griffin} MJ, {Abergel} A, {Abreu} A, {Ade} PAR, {Andr{\'e}} P, {Augueres} JL,
  {Babbedge} T, {Bae} Y, {Baillie} T, {Baluteau} JP, {Barlow} MJ, {Bendo} G,
  {Benielli} D, {Bock} JJ, {Bonhomme} P, {Brisbin} D, {Brockley-Blatt} C,
  {Caldwell} M, {Cara} C, {Castro-Rodriguez} N, {Cerulli} R, {Chanial} P,
  {Chen} S, {Clark} E, {Clements} DL, {Clerc} L, {Coker} J, {Communal} D,
  {Conversi} L, {Cox} P, {Crumb} D, {Cunningham} C, {Daly} F, {Davis} GR, {de
  Antoni} P, {Delderfield} J, {Devin} N, {di Giorgio} A, {Didschuns} I,
  {Dohlen} K, {Donati} M, {Dowell} A, {Dowell} CD, {Duband} L, {Dumaye} L,
  {Emery} RJ, {Ferlet} M, {Ferrand} D, {Fontignie} J, {Fox} M, {Franceschini}
  A, {Frerking} M, {Fulton} T, {Garcia} J, {Gastaud} R, {Gear} WK, {Glenn} J,
  {Goizel} A, {Griffin} DK, {Grundy} T, {Guest} S, {Guillemet} L, {Hargrave}
  PC, {Harwit} M, {Hastings} P, {Hatziminaoglou} E, {Herman} M, {Hinde} B,
  {Hristov} V, {Huang} M, {Imhof} P, {Isaak} KJ, {Israelsson} U, {Ivison} RJ,
  {Jennings} D, {Kiernan} B, {King} KJ, {Lange} AE, {Latter} W, {Laurent} G,
  {Laurent} P, {Leeks} SJ, {Lellouch} E, {Levenson} L, {Li} B, {Li} J,
  {Lilienthal} J, {Lim} T, {Liu} SJ, {Lu} N, {Madden} S, {Mainetti} G,
  {Marliani} P, {McKay} D, {Mercier} K, {Molinari} S, {Morris} H, {Moseley} H,
  {Mulder} J, {Mur} M, {Naylor} DA, {Nguyen} H, {O'Halloran} B, {Oliver} S,
  {Olofsson} G, {Olofsson} HG, {Orfei} R, {Page} MJ, {Pain} I, {Panuzzo} P,
  {Papageorgiou} A, {Parks} G, {Parr-Burman} P, {Pearce} A, {Pearson} C,
  {P{\'e}rez-Fournon} I, {Pinsard} F, {Pisano} G, {Podosek} J, {Pohlen} M,
  {Polehampton} ET, {Pouliquen} D, {Rigopoulou} D, {Rizzo} D, {Roseboom} IG,
  {Roussel} H, {Rowan-Robinson} M, {Rownd} B, {Saraceno} P, {Sauvage} M,
  {Savage} R, {Savini} G, {Sawyer} E, {Scharmberg} C, {Schmitt} D, {Schneider}
  N, {Schulz} B, {Schwartz} A, {Shafer} R, {Shupe} DL, {Sibthorpe} B, {Sidher}
  S, {Smith} A, {Smith} AJ, {Smith} D, {Spencer} L, {Stobie} B, {Sudiwala} R,
  {Sukhatme} K, {Surace} C, {Stevens} JA, {Swinyard} BM, {Trichas} M,
  {Tourette} T, {Triou} H, {Tseng} S, {Tucker} C, {Turner} A, {Vaccari} M,
  {Valtchanov} I, {Vigroux} L, {Virique} E, {Voellmer} G, {Walker} H, {Ward} R,
  {Waskett} T, {Weilert} M, {Wesson} R, {White} GJ, {Whitehouse} N, {Wilson}
  CD, {Winter} B, {Woodcraft} AL, {Wright} GS, {Xu} CK, {Zavagno} A, {Zemcov}
  M, {Zhang} L, {Zonca} E (2010) {The Herschel-SPIRE instrument and its
  in-flight performance}. \aap 518:L3, \doi{10.1051/0004-6361/201014519},
  \eprint{1005.5123}

\bibitem[{{HIFI Observers' Manual}(2011)}]{HIFIHandbook}
{HIFI Observers' Manual} (2011)
  \urlprefix\url{http://herschel.esac.esa.int/Documentation.shtml}

\bibitem[{{Kunszt} et~al(2000){Kunszt}, {Szalay}, {Csabai}, and
  {Thakar}}]{Kunszt2000}
{Kunszt} PZ, {Szalay} AS, {Csabai} I, {Thakar} AR (2000) {The Indexing of the
  SDSS Science Archive}. In: {Manset} N, {Veillet} C, {Crabtree} D (eds)
  Astronomical Data Analysis Software and Systems IX, Astronomical Society of
  the Pacific Conference Series, vol 216, p 141

\bibitem[{{Lemson} and {Springel}(2006)}]{Lemson2006}
{Lemson} G, {Springel} V (2006) {Cosmological Simulations in a Relational
  Database: Modelling and Storing Merger Trees}. In: {Gabriel} C, {Arviset} C,
  {Ponz} D, {Enrique} S (eds) Astronomical Data Analysis Software and Systems
  XV, Astronomical Society of the Pacific Conference Series, vol 351, p 212

\bibitem[{{Lubow} and {Budav{\'a}ri}(2013)}]{Lubow2013}
{Lubow} S, {Budav{\'a}ri} T (2013) {Hubble Source Catalog}. In: {Friedel} DN
  (ed) Astronomical Data Analysis Software and Systems XXII, Astronomical
  Society of the Pacific Conference Series, vol 475, p 203

\bibitem[{{Ott}(2010)}]{Ott2010}
{Ott} S (2010) {The Herschel Data Processing System -- HIPE and Pipelines -- Up
  and Running Since the Start of the Mission}. In: {Mizumoto} Y, {Morita} KI,
  {Ohishi} M (eds) Astronomical Data Analysis Software and Systems XIX,
  Astronomical Society of the Pacific Conference Series, vol 434, p 139,
  \eprint{1011.1209}

\bibitem[{{PACS Observers' Manual}(2013)}]{PACSHandbook}
{PACS Observers' Manual} (2013)
  \urlprefix\url{http://herschel.esac.esa.int/Documentation.shtml}

\bibitem[{{Pilbratt} et~al(2010){Pilbratt}, {Riedinger}, {Passvogel}, {Crone},
  {Doyle}, {Gageur}, {Heras}, {Jewell}, {Metcalfe}, {Ott}, and
  {Schmidt}}]{Herschel}
{Pilbratt} GL, {Riedinger} JR, {Passvogel} T, {Crone} G, {Doyle} D, {Gageur} U,
  {Heras} AM, {Jewell} C, {Metcalfe} L, {Ott} S, {Schmidt} M (2010) {Herschel
  Space Observatory. An ESA facility for far-infrared and submillimetre
  astronomy}. \aap 518:L1, \doi{10.1051/0004-6361/201014759},
  \eprint{1005.5331}

\bibitem[{{Poglitsch} et~al(2010){Poglitsch}, {Waelkens}, {Geis},
  {Feuchtgruber}, {Vandenbussche}, {Rodriguez}, {Krause}, {Renotte}, {van
  Hoof}, {Saraceno}, {Cepa}, {Kerschbaum}, {Agn{\`e}se}, {Ali}, {Altieri},
  {Andreani}, {Augueres}, {Balog}, {Barl}, {Bauer}, {Belbachir}, {Benedettini},
  {Billot}, {Boulade}, {Bischof}, {Blommaert}, {Callut}, {Cara}, {Cerulli},
  {Cesarsky}, {Contursi}, {Creten}, {De Meester}, {Doublier}, {Doumayrou},
  {Duband}, {Exter}, {Genzel}, {Gillis}, {Gr{\"o}zinger}, {Henning},
  {Herreros}, {Huygen}, {Inguscio}, {Jakob}, {Jamar}, {Jean}, {de Jong},
  {Katterloher}, {Kiss}, {Klaas}, {Lemke}, {Lutz}, {Madden}, {Marquet},
  {Martignac}, {Mazy}, {Merken}, {Montfort}, {Morbidelli}, {M{\"u}ller},
  {Nielbock}, {Okumura}, {Orfei}, {Ottensamer}, {Pezzuto}, {Popesso},
  {Putzeys}, {Regibo}, {Reveret}, {Royer}, {Sauvage}, {Schreiber}, {Stegmaier},
  {Schmitt}, {Schubert}, {Sturm}, {Thiel}, {Tofani}, {Vavrek}, {Wetzstein},
  {Wieprecht}, and {Wiezorrek}}]{PACS}
{Poglitsch} A, {Waelkens} C, {Geis} N, {Feuchtgruber} H, {Vandenbussche} B,
  {Rodriguez} L, {Krause} O, {Renotte} E, {van Hoof} C, {Saraceno} P, {Cepa} J,
  {Kerschbaum} F, {Agn{\`e}se} P, {Ali} B, {Altieri} B, {Andreani} P,
  {Augueres} JL, {Balog} Z, {Barl} L, {Bauer} OH, {Belbachir} N, {Benedettini}
  M, {Billot} N, {Boulade} O, {Bischof} H, {Blommaert} J, {Callut} E, {Cara} C,
  {Cerulli} R, {Cesarsky} D, {Contursi} A, {Creten} Y, {De Meester} W,
  {Doublier} V, {Doumayrou} E, {Duband} L, {Exter} K, {Genzel} R, {Gillis} JM,
  {Gr{\"o}zinger} U, {Henning} T, {Herreros} J, {Huygen} R, {Inguscio} M,
  {Jakob} G, {Jamar} C, {Jean} C, {de Jong} J, {Katterloher} R, {Kiss} C,
  {Klaas} U, {Lemke} D, {Lutz} D, {Madden} S, {Marquet} B, {Martignac} J,
  {Mazy} A, {Merken} P, {Montfort} F, {Morbidelli} L, {M{\"u}ller} T,
  {Nielbock} M, {Okumura} K, {Orfei} R, {Ottensamer} R, {Pezzuto} S, {Popesso}
  P, {Putzeys} J, {Regibo} S, {Reveret} V, {Royer} P, {Sauvage} M, {Schreiber}
  J, {Stegmaier} J, {Schmitt} D, {Schubert} J, {Sturm} E, {Thiel} M, {Tofani}
  G, {Vavrek} R, {Wetzstein} M, {Wieprecht} E, {Wiezorrek} E (2010) {The
  Photodetector Array Camera and Spectrometer (PACS) on the Herschel Space
  Observatory}. \aap 518:L2, \doi{10.1051/0004-6361/201014535},
  \eprint{1005.1487}

\bibitem[{{SPIRE Handbook}(2014)}]{SPIREHandbook}
{SPIRE Handbook} (2014)
  \urlprefix\url{http://herschel.esac.esa.int/Documentation.shtml}

\bibitem[{{Szalay} et~al(2002){Szalay}, {Gray}, {Thakar}, {Kunszt}, {Malik},
  {Raddick}, {Stoughton}, and {vandenBerg}}]{Szalay2002}
{Szalay} AS, {Gray} J, {Thakar} AR, {Kunszt} PZ, {Malik} T, {Raddick} J,
  {Stoughton} C, {vandenBerg} J (2002) {The SDSS SkyServer: Public Access to
  the Sloan Digital Sky Server Data}. eprint arXiv:cs/0202013
  \eprint{cs/0202013}

\end{thebibliography}

\end{document}